\documentclass{aa}  
\usepackage{graphicx}
\usepackage{txfonts}
\usepackage{comment}
\usepackage{siunitx}
\usepackage{hyperref}
\usepackage{lscape}
\usepackage{longtable}
\usepackage{csquotes}
\usepackage{orcidlink}
\usepackage{natbib}
\bibpunct{(}{)}{;}{a}{}{,} 

\usepackage{placeins}
\graphicspath{{figs/}}

\newcommand{\kms}[1]{\SI{#1}{\kilo\metre\per\second}}
\newcommand{\hcl}{H$^{35}$Cl\xspace} 
\newcommand{\hcliso}{H$^{37}$Cl\xspace} 
\newcommand{\htclp}{H$_{2}$Cl$^+$\xspace} 
\newcommand{\hclp}{HCl$^+$\xspace} 

\newcommand{\voff}{$\upsilon_\mathrm{off}$\xspace} 
\newcommand{\vLSR}{$\upsilon_\mathrm{LSR}$\xspace} 
\newcommand{\Dv}{$\Delta\upsilon$\xspace}

\newcommand{\Ntot}{$N_\mathrm{tot}$\xspace} 

\newcommand{\mychange}[1]{\textcolor{black}{#1}}
\newcommand{\mychangere}[1]{\textcolor{black}{#1}}

\begin{document}

   \title{APEX survey of interstellar HCl} 
   \titlerunning{APEX survey of interstellar HCl} 
   \subtitle{$^{35}\text{Cl} / ^{37}\text{Cl}$ isotopic ratios in dense cores and outflows}

   \author{Lennart M. B\"{o}hm\orcidlink{0009-0002-4834-9247}\inst{1, 2}
          \and
          Arshia M. Jacob\orcidlink{0000-0001-7838-3425}\inst{2, 3}
          \and
          Friedrich Wyrowski\inst{2}
          \and
          Karl M. Menten\inst{2}\thanks{Deceased}
          \and
          Katharina Immer\orcidlink{0000-0003-4140-5138}\inst{1}
          \and
          Ashley T. Barnes\orcidlink{0000-0003-0410-4504}\inst{1}
          }
    \authorrunning{B\"{o}hm et al.
          }

    \institute{
    European Southern Observatory, Karl-Schwarzschild-Straße 2, 85748 Garching bei München, Germany
    \and
    Max-Planck-Institut f\"{u}r Radioastronomie, Auf dem H\"{u}gel 69, 53121 Bonn, Germany
    \and 
    I. Physikalisches Institut, Universit\"at zu K\"oln, Z\"ulpicher Str. 77, 50937 K\"oln, Germany
      }

   \date{Received Aug 8, 2025; accepted Nov 25, 2025}

    \abstract
    {Despite being only the {19th} most abundant element in the interstellar medium, chlorine’s reactivity and volatility give rise to a unique interstellar chemistry, favouring the formation of several chlorine-bearing hydrides. 
    Further, the $^{35}\text{Cl} / ^{37}\text{Cl}$ ratio -- shaped in supernovae and evolved stars -- probes nucleosynthesis across the Galaxy. 
    Yet, studies of Cl-bearing molecules have remained limited to a few sightlines due to observational challenges.
    }
    {We systematically investigated the Galactic distribution of HCl and the [H$^{35}$Cl]/[H$^{37}$Cl] ratio in high-mass star-forming regions. As a probe of a region's nucleosynthesis history, this ratio may constrain predictions of Galactic chemical evolution models.
    }
    {We surveyed the ground-state $J=1$$-$0 transitions of \hcl and \hcliso near $\SI{625}{\giga\hertz}$ toward $28$ sources with the SEPIA660 receiver on the APEX $12\,$m sub-millimetre telescope. This survey more than doubles the number of sources with HCl detections and reveals HCl emission arising from both the background core and associated outflows. The spectra were modelled with XCLASS to derive column densities, isotopic ratios, and the kinematics of both the core and the outflow components.
    }
    {\hcl was detected in all sources, \hcliso in all but two, with spectral line profiles ranging from those with only emission to complex emission–absorption mixtures. Column densities span from $2.3$$-$$22.8\times 10^{13}\,\mathrm{cm}^{-2}$ for \hcl and $0.6$$-$$12.5\times 10^{13}\,\mathrm{cm}^{-2}$ for \hcliso, resulting in isotopic ratios between $1.6$ and $3.5$ in emission-only sources.}
    {
    The derived [\hcl\!\!]$/$[\hcliso\!\!] aligns with Galactic chemical evolution models and shows no trend with Galactocentric radius. However, local variations may reflect recent nucleosynthesis. Overall, the results suggest that most Galactic chlorine was synthesized during epochs of lower average metallicity in the Galaxy. Notably, we detect \hcl emission arising from outflows -- particularly explosive ones -- hinting at its presence in a broader range of environments. The present single-dish observations cannot reveal the origin of HCl in outflows; necessitating interferometric follow-up observations.
    }


   \maketitle

\section{Introduction}
\label{sec:intro}
\label{sec:intro-chemistry-and-isotopic-ratio}
Massive stars with initial masses $\geq$$8\,M_{\odot}$ and their explosive deaths as supernovae (SNe) are not only important sources of kinematic energy in the interstellar medium (ISM) and molecular clouds, but also influence subsequent star-formation through feedback processes and replenish material into the ISM  \citep{motte2018high, rosen2020zooming}. 
In particular, they are the primary sites of heavy element formation, imprinting their nucleosynthetic signatures onto the material they enrich. Among these, halogens like chlorine have been of particular interest to astronomers owing to their volatile nature and simple chemistry despite their low elemental abundances, e.g., $[\mathrm{Cl}]/[\mathrm{H}] = 1.8\times10^{-7}$\citep{neufeld2009chemistry,gerin2016interstellar}. 

\begin{figure}
    \centering
    \includegraphics[width=0.4\textwidth]{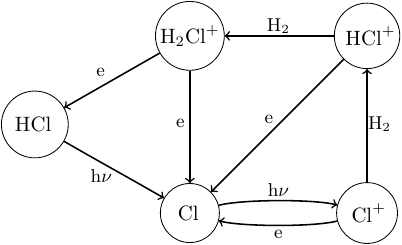}
    \caption[Chemical network of Cl-bearing species with the dominant reactions in the ISM]{Chemical network displaying the most relevant gas-phase formation and destruction pathways involved in interstellar chlorine chemistry adapted from Fig.~3 of \citet{neufeld2021chemistry}.}
    \label{fig:cl-network}
\end{figure}

Present in the ISM predominantly as Cl$^+$, due to its ionisation potential (IP$_{\rm Cl} = \SI{12.97}{eV}$) being slightly lower than that of hydrogen (IP$_{\rm H} = \SI{13.60}{eV}$), Cl$^+$ is one of only two species, alongside fluorine, that can react exothermically with molecular hydrogen in the diffuse ISM. This hydrogen abstraction reaction forms HCl$^+$ and subsequently initiates interstellar chlorine chemistry. 
Figure~\ref{fig:cl-network} illustrates the chemical network for the gas-phase production of chlorine hydrides and hydride ions (HCl, \hclp, \htclp), reduced to the dominant formation and destruction reactions for each species. Of these chlorine-bearing hydrides, HCl has been the most widely observed and is commonly used as a proxy for determining the chlorine abundance in stars and the ISM, via its rotational-vibrational transition at $\SI{3.7}{\micro\metre}$ \citep{maas2016rovibHCl}, its electronic transition at $\SI{1290}{\angstrom}$ \citep{federman1995vibrationally} and its ground-state rotational lines near $\SI{625}{\giga\hertz}$ \citep{pickett1998jpl}, respectively. 

Moreover, chlorine exists in two stable isotopes, $^{35}$Cl and $^{37}$Cl, both of which are predominantly formed in core-collapse supernovae (CCSNe) with their relative abundances also being influenced by the explosive burning of oxygen in asymptotic giant branch (AGB) stars via fast reactions with the more abundant elements $^{34}$S and $^{36}$Ar, respectively \citep{gerin2016interstellar}. 
Further, \citet{pignatari2010weak} used new neutron capture cross-sections to refine the abundance of $^{35}$Cl and $^{37}$Cl produced from the weak s-process in massive stars -- a production channel which potentially alters local isotopic ratios. 
CCSNe and AGB star models, summarised in \citet{maas2018chlorine}, predict isotopic ratios \mychange{in between $1.2$ and $4.2$}, depending on physical parameters including the progenitor star's initial mass, and the metallicity and energy of the CCSNe. For instance, in contrast to lower-metallicity ($Z=0.014$) CCSNe, the chemical enrichment in higher-metallicity ($Z=0.02$) CCSNe may enhance the production of $^{37}$Cl relative to $^{35}$Cl. The same trend is seen for more massive progenitor stars ($25\,M_\odot$, compared to $13\,M_\odot$) leading to $^{35}$Cl/$^{37}$Cl isotopic ratios lower than the value of $3.13$\mychange{, measured in the Solar System \citep[by infrared spectroscopy of HCl in Sun spots; see Table 10 of ][]{lodders20094}}. Therefore, investigating variations in the $^{35}$Cl/$^{37}$Cl isotopic ratio across the Galaxy can potentially serve as a tracer of the nucleosynthesis history in massive star-forming regions.
However, observations of the fundamental rotational transitions of HCl ($J=1$--0) near $\SI{625.9}{\giga\hertz}$ are challenging due to its proximity to an atmospheric water feature at $\SI{621}{\giga\hertz}$, resulting in low atmospheric transmission. 
As a consequence, to date, HCl has only been detected toward a limited number of sight lines.

Observationally, while \citet{blake1985chlorine} reported the first detection of \hcl in the ISM using the Kuiper Airborne Observatory (KAO) toward OMC-1, the detection of both isotopes of HCl was made only a decade later by \citet{salez1996hydrogen} using the Caltech Submillimeter Observatory (CSO). These authors reported a $[\mathrm{H^{35}Cl}]/[\mathrm{H^{37}Cl}]$ isotopic ratio of $4$ to $6$ toward Orion~A, providing the first observational evidence for deviations in chlorine's isotopic ratio from \mychange{the Solar System value of $3.13$ \citep{lodders20094}}. 
Critically, these observations were unable to resolve the hyperfine-structure splitting (HFS) of HCl's $J=1$--0 transition consisting of three components separated by less than $\SI{9}{\kilo\metre\per\second}$, due to limited sensitivity and spectral resolution. It was only with the high spectral resolution provided by the Heterodyne Instrument for Far-Infrared (HIFI) receiver aboard the Herschel Space Observatory (HSO), that the HFS lines of HCl were resolved \citep{cernicharo201035cl}. 
Soon after, using the CSO, \citet{peng2010comprehensive} presented the first comprehensive study of HCl toward 27 Galactic sources probing a range of environments from evolved stars and giant molecular clouds to star-forming regions and supernova remnants (SNRs). Toward a subset of their sample, predominantly composed of sources associated with star-forming regions, these authors also reported the detection of \hcliso. 
However, owing to the complex nature of the observed spectral line profiles, these authors were only able to report isotopic ratios toward 11 sources. The 
[$\mathrm{H^{35}Cl}]/[\mathrm{H^{37}Cl}$] isotopic ratios they determined range from $0.9$ to $4.2$ with an average integrated intensity ratio of $2.5\pm1.1$, in sources covering galactocentric distances ranging from 5.2~to $\SI{9.6}{kpc}$ \citep[][]{reid2019trigonometric}.
Overall, observations of \mychange{HCl isotopologues} have largely been restricted to only a handful of Galactic sight lines, 
in part due to limitations in the sensitivity and spectral resolution achievable, as well as the limited operating time of air-/space-borne missions. \\

Taking advantage of its access to high-frequencies, excellent weather conditions, and advanced receiver technology enabling high spectral resolution observations, we used the Atacama Pathfinder Experiment (APEX) $\SI{12}{\metre}$ sub-millimetre (sub-mm) telescope\footnote{The data was collected under the Atacama Pathfinder
EXperiment (APEX) Project, led by the Max Planck Institute for Radio Astronomy (MPIfR)
hosted and operated by ESO on behalf of the MPIfR.} to carry out a comprehensive survey of \hcl and \hcliso across the Galaxy. 
Building on previous observations, this study extends the analysis of chlorine isotopic ratios to galactocentric radii between $0.1$ and $\sim$$\SI{10}{\kilo pc}$, with G5.89$-$0.39 at $\SI{5.2}{kpc}$ \citep{reid2019trigonometric} being the innermost Milky Way source where this ratio has been previously observed \citep{peng2010comprehensive}.
Sections~\ref{sec:observations}, and \ref{sec:results} outline the observing strategy and present the spectral line analysis carried out while we discuss in Sect.~\ref{sec:discussion} variations in the derived chlorine isotopic ratio across Galactic scales, its implications for Galactic chemical evolution models, and unexpected kinematics of HCl in outflows. Lastly, we summarise our findings in Sect.~\ref{sec:conclusions}.

\begin{figure}
    \centering
    \includegraphics[width=0.49\textwidth]{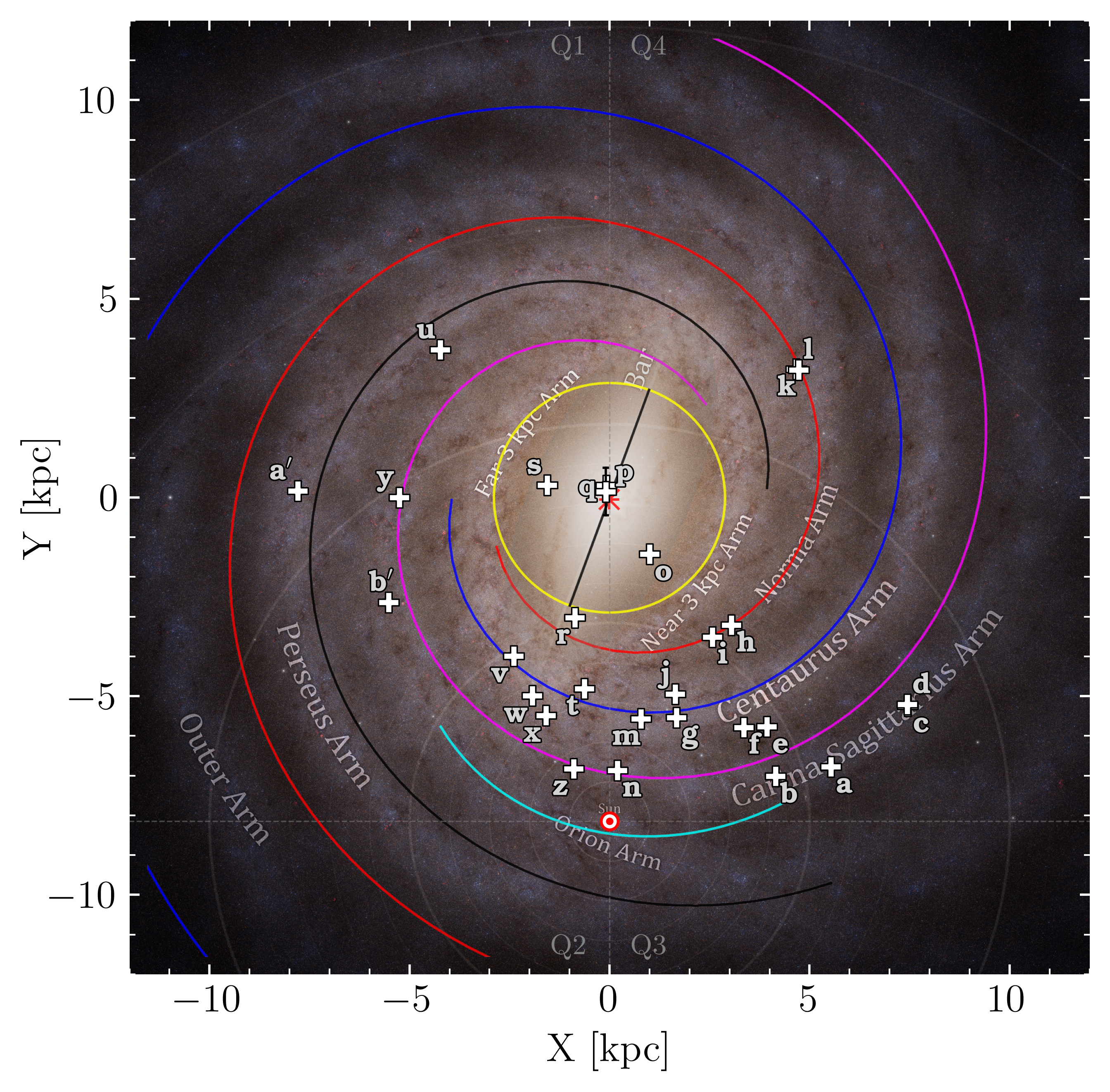}
    \caption[Source distribution across the Milky Way]{Distribution of the targets observed in this work across the Milky Way in white \enquote{$+$} symbols along with their identifiers, as listed in Table~\ref{tab:source-params}. 
    The red asterisk and Sun symbols mark the Galactic centre at (0, 0)\,kpc and the position of the Solar System at (0, $-8.15$)\,kpc, respectively \mychange{\citep{reid2019trigo_distances}}. 
    The sources are projected onto the Galactic plane with parallax-based distances from \citet{reid2019trigonometric}.
    The underlying Milky Way spiral arm pattern follows the log-periodic model from the same study, with colours adapted from \citet{jacob2022hygal}.
    The background image, is an artist's impression of the large-scale structure of the Milky Way, courtesy of ESA/Gaia/DPAC/Stefan Payne-Wardenaar. 
    }
    \label{fig:source-distribution}
\end{figure}

\section{Observations}
\label{sec:observations}
\begin{table*}[htb]
    \centering
    \caption[Observed targets in this survey]{List of observed sources and their properties.} 
    \small
    \begin{tabular}{clccrrrr}
           \hline\hline
        \# & Source & \multicolumn{1}{c}{Right Ascension} & \multicolumn{1}{c}{Declination} &  \multicolumn{1}{c}{$\upsilon_\mathrm{LSR}$}& \multicolumn{1}{c}{$d$} & \multicolumn{1}{c}{$R_\mathrm{gal}$} & \multicolumn{1}{c}{$T_\mathrm{c}$} \\
        &  Designation & \multicolumn{1}{c}{[hh:mm:ss]} & \multicolumn{1}{c}{[dd:mm:ss]}& \multicolumn{1}{c}{[$\mathrm{km\,s}^{-1}$]}& \multicolumn{1}{c}{[kpc]}   & \multicolumn{1}{c}{[kpc]} & \multicolumn{1}{c}{[K]} \\ \hline
        \multicolumn{7}{c}{Galactic Sources} \\
        \hline
        a & HGAL~284.015--00.857 & 10:20:16.1 & $-$58:03:55.0 & 10.4 &  5.7 &  9.0 & 0.6 \\
        b  &  HGAL~285.264--00.049 & 10:31:29.5 & $-$58:02:19.5 & 3.4 &  4.3  &  8.2 & 1.5 \\
        c & G291.272--00.714 & 11:11:51.6 & $-$61:18:39.5 & $-$22.8$^*$\negmedspace\! & 8.0 & $\sim 9.3$ & 0.6 \\
        d & G291.579--00.431 & 11:15:05.7 & $-$61:09:40.8 & 12.8 &  8.0  &  9.3 & 1.1 \\
        e & IRAS~12326--6245 & 12:35:35.9 & $-$63:02:29.0 & $-$38.6 &  4.6  &  7.2 & 3.8 \\
        f & G305.209+00.206 & 13:11:14.2 & $-$62:34:42.1 & $-$42.2$^*$\negmedspace\! & 4.1  & 8.0 & 1.8 \\
        g & G327.293--00.579 & 15:53:07.8 & $-$54:37:06.4 & $-$44.9 & 3.1 & 6.2 & 7.8 \\
        h & G328.307+00.423 & 15:54:07.2 & $-$53:11:40.0 & $-$91.1 &  5.8  &  4.6 & 1.4 \\
        i & IRAS~16060--5146 (G330.95) & 16:09:53.0 & $-$51:54:54.8 & $-$92.2 & 5.3 & 4.5 & 8.3 \\
        j & IRAS~16164--5046 (G332.83) & 16:20:10.6 & $-$50:53:17.6 & $-$57.4$^*$\negmedspace\!      & 3.6 & 5.4 & 5.6 \\
        k & G337.704--00.054 & 16:38:29.4 & $-$47:00:38.8 & $-$48.6 & 12.3  & 5.1 & 2.5 \\
        l & IRAS~16352--4721 & 16:38:50.6 & $-$47:28:04.0 & $-$40.2 &  12.3  &  5.1 & 3.2 \\
        m & IRAS~16547--4247 & 16:58:17.2 & $-$42:52:08.9 & $-$32.7 &  2.7  &  5.8 & 3.5 \\
        n & NGC~6334I & 17:20:53.4 & $-$35:47:01.5 & $-$7.5 &  1.3  &  7.0  & 11.8 \\
        o & G351.581--00.352 & 17:25:25.0 & $-$36:12:45.3 & $-$96.7 &  6.8  & 1.7 & 5.1 \\
        p & Sgr~B2~(M) & 17:47:20.2 & $-$28:23:05.0 & 63.0$^*$\negmedspace\! & 8.3  & 0.1 & 19.4 \\
        q & HGAL~000.546--00.852 & 17:50:14.5 & $-$28:54:30.7 & 16.8 &  7.7–9.2   &  0.4–1.0 & 3.7 \\
        r & G009.621+00.194 & 18:06:14.9 & $-$20:31:37.0 & 4.2 &  5.2   &  3.3 & 2.0 \\
        s & G010.472+00.027 & 18:08:38.2 & $-$19:51:49.6 & 68.0 &  8.6  & 1.6 & 5.6 \\
        t & W31~C (G010.624--00.384) & 18:10:28.7 & $-$19:55:49.7 & $-$0.7 & 3.4  & 7.1 & 5.5 \\
        u & G019.609-00.234 & 18:27:38.0 & $-$11:56:36.6 & 40.0 &  12.6  & 4.7 & 3.3 \\
        v & G029.954--00.016 & 18:46:03.7 & $-$02:39:21.2 & 97.0 &  4.8 &  4.7 & 2.1 \\
        w & G031.412+00.307 & 18:47:34.3 & $-$01:12:46.0 & 97.4 & 3.7  & 5.4 & 3.9 \\
        x & W43-MM1 & 18:47:47.0 & $-$01:54:28.0 & 96.4 &  3.1 &  5.7  & 4.2 \\
        y & G032.797+00.191 & 18:50:30.6 & $-$00:02:00.0 & 14.9 &  9.7   &  5.3 & 2.0 \\
        z & G034.258+00.154 & 18:53:18.7 & +01:14:58.0 & 56.8 &  1.6  &  7.0 & 8.0 \\
        a$^\prime$ & W49~N & 19:10:13.2 & +09:06:12.0 & 4.9$^*$\negmedspace\! & 11.4  & 7.8 & 8.0 \\
        b$^\prime$ & G045.071+00.132 & 19:13:22.0 &  +10:50:54.0 & 59.7 &  7.8  &  6.1 & 1.3 \\ \hline

           \hline
        \end{tabular}

        \tablefoot{From left to right the columns are: the source identifier, the source designation, the source coordinates (J2000), their systemic velocities (based on that observed for the H$^{13}$CO$^+$\,(7--6) line), their heliocentric and galactocentric distances, and the observed continuum temperature \mychange{(i.e., the average main beam temperature of line-free channels)} around the \hcl\,($J = 1$--0) line. The labels in the first column are used in Fig.~\ref{fig:source-distribution}. Systemic velocities that are marked with an asterisk are taken from \citet{giannetti2014atlasgal} using C$^{17}$O while that of G{291.272--00.714} is taken from \citet{urquhart2007rms} and determined using $^{13}$CO.}
        \tablebib{The Galactic heliocentric distances have been taken from \citet{jacob2022hygal} and references therein and cross-checked with version~2 of the parallax-based distance calculator on \href{http://bessel.vlbi-astrometry.org/node/378}{http://bessel.vlbi-astrometry.org/node/378} \citep{reid2019trigonometric}. 
        }
        \label{tab:source-params}
\end{table*}
The single-pointing observations presented here were carried out under project ID M9523C\_109 (PI: Karl M. Menten) on June 17--19 and July 26, 2022, using the Swedish-ESO PI Instrument \citep[SEPIA660;][]{belitsky2018sepia}, on the APEX $\SI{12}{\metre}$ sub-mm telescope \citep{gusten2006atacama}. The sources targeted in this survey were selected from the Hi-GAL Herschel Key-guaranteed time project catalogue \citep{molinari2010hi, elia2021hi}, with continuum fluxes at $\SI{500}{\micro\metre}$ (close to the HCl\,(1--0) transitions at $\sim$$\SI{480}{\micro\metre}$) in excess of $\sim$$\SI{100}{Jy}$. The final source sample comprises of 28 Galactic star-forming regions across a wide range of galactocentric radii between $0.1$ and $\sim$$\SI{10}{kpc}$, allowing us to sample HCl in a variety of different Galactic environments. 
The distribution of the sources across the Milky Way is displayed in Fig.~\ref{fig:source-distribution}. 
The sources, along with their coordinates, the observed local standard of rest \mychange{(LSR)} velocity, \vLSR, and continuum temperature \mychange{(i.e., the average main beam temperature of line-free channels)} at $\SI{625}{\giga\hertz}$ are summarised in Table~\ref{tab:source-params}.

\begin{figure}
    \centering
    \includegraphics[width=0.49\textwidth]{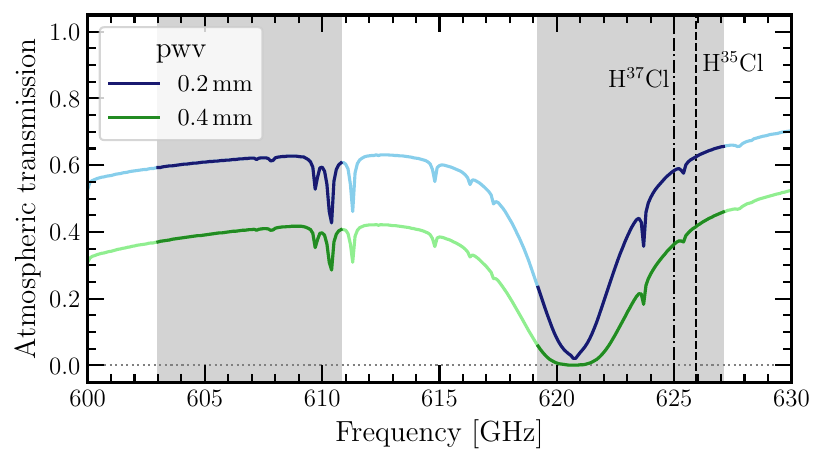}
    \caption[The atmospheric transmission in the observed frequency range at the telescope site]{ Atmospheric transmission at the APEX $\SI{12}{\metre}$ sub-mm telescope site in the frequency range from $\SIrange{600}{630}{\giga\hertz}$ for typical values of pwv columns in our observations: 
    $\SI{0.2}{\milli\metre}$ (blue) and $\SI{0.4}{\milli\metre}$ (green). 
    The dashed and dash-dotted black lines correspond to the $J = 1$$-$$0$ transition of \hcl and \hcliso, respectively, with the grey shaded regions marking the sideband coverage of our observations. 
    }
    \label{fig:atm-trans}
\end{figure}

For our observations, the SEPIA660 dual polarisation, sideband separated heterodyne receiver was tuned to $\SI{609}{\giga\hertz}$ in the lower sideband 
and covered a bandwidth of $\SI{7.9}{\giga\hertz}$ in each sideband. This resulted in a frequency coverage ranging from $\SIrange{602.9}{610.8}{\giga\hertz}$ in the lower sideband and $\SIrange{619.2}{627.1}{\giga\hertz}$ in the upper sideband. The frequencies and spectroscopic properties of the HCl lines studied in this work are summarised in Table~\ref{tab:spectral-lines}.
The observations were conducted under excellent weather conditions, with precipitable water vapour (pwv) levels between $0.2$ and $\SI{0.4}{\milli\metre}$. To further emphasise the importance of these conditions for observing HCl, Fig.~\ref{fig:atm-trans} shows the atmospheric transmission at the APEX site. 
To ensure reliable baselines, the observations were carried out in wobbler switching mode with a wobbling rate of $\SI{1.5}{\hertz}$ and a maximum symmetric throw of $60^{\prime\prime}$ in azimuth. The total ON-time was $\SI{7}{\minute}$ per target.

An advanced version of the \mychange{eXtended bandwidth Fast Fourier Transform Spectrometer} (XFFTS) backend developed at the \mychange{Max Planck Institute for Radio Astronomy} (MPIfR) \citep{klein2012high} provided a spectral resolution of $\SI{61}{\kilo\hertz}$ corresponding to a velocity resolution of $\SI{0.03}{\kilo\metre\per\second}$. The half power beam width (HPBW) at $\SI{625}{\giga\hertz}$ is $10\rlap{.}^{\prime\prime}1$.
\begin{table*}[htb]
    \centering
    \caption[Spectral line properties of HCl rotational ground state transitions]{Spectroscopic properties  
    of the $J=1$--0 lines of \hcl and \hcliso. %
    }
    \label{tab:spectral-lines}
    \begin{tabular}{ccclcccc}
    \hline\hline
    \multicolumn{1}{c}{Species} & \multicolumn{2}{c}{Transition} & \multicolumn{1}{c}{Frequency} & \multicolumn{1}{c}{HFS velocity offset} & \multicolumn{1}{c}{$E_\mathrm{up}$} & \multicolumn{1}{c}{$g_\mathrm{up}$} & \multicolumn{1}{c}{$A_{ij}$} \\
    \multicolumn{1}{c}{} & \multicolumn{1}{c}{$J^{\prime}\,$ -- $J^{\prime\prime}$} & \multicolumn{1}{c}{$F^{\prime}\,$ -- $F^{\prime\prime}$} & \multicolumn{1}{c}{[GHz]} & \multicolumn{1}{c}{[$\mathrm{km\,s^{-1}}$]} & \multicolumn{1}{c}{[K]} & \multicolumn{1}{c}{} & \multicolumn{1}{c}{$\times10^{-3}$\,[s$^{-1}$]} \\ \hline
    \hcliso & $1$ -- $0$ & $3/2$ -- $3/2$ & 624.964374(100) & \hspace{0.65em}$6.45$ & 30.0  & 4 & 1.19258 \\
       & & $5/2$ -- $3/2$ & 624.977821(100)$^*$ & \hspace{0.65em}$0.00$ & & 6 & 1.19263 \\
       & & $1/2$ -- $3/2$ & 624.988334(100) & $-5.04$ &  & 2 & 1.19271 \\
    \hcl & $1$ -- $0$ & $3/2$ -- $3/2$  & 625.901603(100) & \hspace{0.65em}$8.22$ & 30.0  & 4 & 1.16915 \\
       & & $5/2 $ -- $3/2$ & 625.918756(100)$^*$ & \hspace{0.65em}$0.00$ & & 6 & 1.16920  \\
       & & $1/2$ -- $3/2$ & 625.932007(100) & $-6.35$ & & 2 & 1.16929  \\ \hline
    \end{tabular}   
    \tablefoot{The frequencies marked with an asterisk correspond to the strongest HFS line and is used as the rest frequency to set the velocity scale. \mychange{Numbers in parentheses indicate experimental errors in the frequency in the last decimal place.}}
    \tablebib{The spectroscopic parameters are taken from the JPL spectral line catalogue \citep{pickett1998jpl} with experimental measurements carried out by \citet{deLeluw1973} and \citet{Nolt1987}.}
\end{table*}

It should be noted that the real-time calibration process accounts for variations in transmission near the atmospheric absorption features, originating mostly from water vapour transitions. This is achieved through a channel-by-channel atmospheric calibration, where each frequency channel is individually adjusted to correct for atmospheric effects. 
Post calibration, the subsequent data analysis was carried out in Python using the GILDAS-CLASS\footnote{\href{https://www.iram.fr/IRAMFR/GILDAS}{https://www.iram.fr/IRAMFR/GILDAS}} FITS format data files \citep{pety2005gildas} converted to ASCII files. 
The data were converted from antenna temperature scale, $T_\text{A}^*$, to main beam temperature scale, $T_\text{mb}$, using a forward efficiency, $\eta_{\rm f}$, of $0.95$ and a main beam efficiency, $\eta_\text{mb}$, of $0.49$ (determined from observations of Mars\footnote{See \href{https://www.apex-telescope.org/telescope/efficiency/?yearBy=2022}{https://www.apex-telescope.org/telescope/efficiency/?yearBy=2022} for more information.}). The spectra were smoothed to $\SI{0.51}{\kilo\metre\per\second}$ using a boxcar averaging method.
 
The baselines were fit with polynomial orders between 0 and 3, with the best-fit order determined 
based on the Anderson-Darling (AD) test statistic, implemented through the \texttt{anderson} function from the \texttt{scipy.stats} module \citep{gommers2022scipy} in Python.
The AD test is more sensitive to deviations in the tails of the distribution (in this case, the line-free channels in a given spectrum) compared to the more general Kolmogorov-Smirnov test. 
The line-free channels are selected for each source individually and a similar number of channels is guaranteed on both sides of the HCl lines.
However, toward NGC\,6334I fitting a polynomial baseline was not straightforward. 
This is due to the spectral richness, often characterised by very broad lines ($\sim$\kms{10}) towards this source, which has resulted in the absence of line-free channels around the \hcl and \hcliso\,(1--0) lines. \mychange{The HCl lines observed towards NGC\,6334I were therefore excluded from further analysis.}

\section{Results and analysis}
\label{sec:results}
The observations presented in this study double the number of targets towards which \hcl and \hcliso have been detected.
A variety of spectral line profiles 
were detected, with 11 sources showing only emission, while the remaining sources display more complex spectral line profiles. 
The general fitting procedure, carried out using the eXtended CASA Line Analysis Software Suite \citep[XCLASS;][]{moller2017extended}, is described in Sect.~\ref{sec:spec-line-analysis-xclass}, followed by a detailed description of the distinct components in Section~\ref{sec:spectral-line-profiles}. 

\subsection{Spectral line fitting}
\label{sec:spec-line-analysis-xclass}
To model the observed line profiles, XCLASS solves the radiative transfer equation for a one-dimensional isothermal source assuming conditions of local thermodynamic equilibrium (LTE).
The lines are defined by a set of input parameters in the \texttt{molfit} file, which includes the source size, $\theta_\mathrm{s}$, the rotational temperature, $T_\mathrm{rot}$, the total column density of the molecule, $N_\mathrm{tot}$, the line width, $\Delta \upsilon$, and the velocity offset, $\upsilon_\mathrm{off}$. While $T_\mathrm{rot}$, $N_\mathrm{tot}$ and $\Delta \upsilon$ define the shape of the line, $\upsilon_\mathrm{off}$ defines the offset of the observed spectral line component with respect to \vLSR for a given source.
Furthermore, from the specified telescope size and observed frequency, XCLASS computes the beam size 
and together with $\theta_\mathrm{s}$, the program accounts for effects of beam dilution. The simple geometry of the code further allows us to specify two types of components: core and foreground, depending on whether the observed spectral line component is considered to arise from the core (background source) or from material in the foreground of the source being modelled. Multiple components are used to fit the observed HCl spectra with most fits including a core and foreground component. 
XCLASS incorporates the latest spectroscopic parameters, taken from the Cologne Database for Molecular Spectroscopy \citep[CDMS;][]{muller2005cologne, 2016JMoSp.327...95E} and Jet Propulsion Laboratory molecular spectroscopy database \citep[JPL;][]{pickett1998jpl}.

In this work, the spectral line fitting is optimised by using a chain of algorithms to explore $\chi^2$ minimisation in a larger parameter space. 
In the first step, the input parameters in the \texttt{molfit} file are optimised using a genetic algorithm, which searches the parameter space for a global minimum. This approach provides better initial parameters which helps avoid local minima, and improves the convergence of the subsequent Levenberg-Marquardt (LM) algorithm. The LM algorithm then minimises the sum of the residuals between the modelled and observed spectra, ultimately yielding an optimised set of parameters. 
While the output of the LM algorithm is considered the optimal fit, a Markov Chain Monte Carlo (MCMC) algorithm is employed in the final step to estimate the errors in the optimal fit parameters. After 200 iterations, the absolute errors are determined by the lower and upper boundaries of the 2$\sigma$ interval around the mode of the resulting parameter distribution, assuming a Gaussian distribution. From the best fit parameters a relative error is determined which is then divided by a factor of 2 to obtain the 1$\sigma$ upper and lower errors.

\subsubsection{Assumptions and uncertainties}
\mychange{We confirmed that the $10\rlap{.}^{\prime\prime}1$ APEX beam was fully filled toward all sources by cross-checking the spatial extent of their dust continuum emission at $\SI{500}{\micro\metre}$ and $\SI{870}{\micro\metre}$. These maps were obtained from the Hi-GAL \citep{elia2021hi} and ATLASGAL\footnote{See \href{https://atlasgal.mpifr-bonn.mpg.de/.}{https://atlasgal.mpifr-bonn.mpg.de/}} \citep{schuller2009atlasgal} catalogues, respectively. }
Consequently, a beam filling factor of $1$ is implemented by fixing the source size parameter to $20^{\prime\prime}$ -- an arbitrary value chosen to exceed the beam size. 

\mychangere{Furthermore, assuming LTE and that the HCl emission arises from dense gas ($n(\mathrm{H{_2}})\approx\!10^5\!-\!10^7\,\mathrm{cm^{-3}}$), we fix $T_\mathrm{rot}$ to the dust temperature, $T_\mathrm{dust}$, for components seen in emission and to $\SI{2.73}{\kelvin}$ for components seen in absorption. 
This density range is not measured directly, but is guided by previous non-LTE studies of HCl in massive star-forming regions \citep{peng2010comprehensive} and by predictions from chlorine chemistry models \citep{neufeld2009chemistry}, which indicate that HCl traces very dense gas. 
At such densities, gas-dust collisions efficiently couple the gas and dust temperatures, so that $T_\mathrm{gas}$ is expected to be close to $T_\mathrm{dust}$ and, under LTE, $T_\mathrm{rot}$ then approximates $T_\mathrm{gas}$, justifying the assumption $T_\mathrm{rot}=T_\mathrm{dust}$ for the dense HCl-emitting component. }
The $T_\mathrm{dust}$ values are obtained from the ATLASGAL catalogue \citep{schuller2009atlasgal} for all sources, except for {HGAL\,284.015$-$00.857}, where it was fixed at $\SI{35}{\kelvin}$, a typical $T_\mathrm{dust}$ value for our sample. We note that while the assumption of $T_{\rm rot}$ = $T_{\rm dust}$ is valid toward cold\mychange{, dense} cores, it may not be true for warmer\mychange{, less dense} outflow or envelope components, therefore, resulting in an overestimation of the column densities toward this component. 
\mychange{The error introduced by fixing $T_\mathrm{rot}$ is discussed in detail in Appendix~\ref{sec:appendix-fixing-Trot}.}
Since we are interested in the ratio between the two isotopologues, the impact of this overestimation cancels out, as we expect both HCl lines to originate from the same gas layers. This assumption of co-spatiality allows us to fit both isotopologues simultaneously in XCLASS with the same \Dv and \voff parameters, while treating the column density of \hcl and thus their isotopic ratio as free parameters. 
Note that the code was modified to ensure that varied components (i.e., arising from the core or outflow and to account for whether the lines are seen in emission or absorption) are not characterised by the same isotopic ratios.  
This approach not only allowed for more stable fits, but also for the optimisation of the isotopic ratio as an independent fit parameter.

Since the assumption of LTE may not be valid for all components in the fit, the derived column densities could be overestimated, especially for more complex spectra. However, this overestimation is less significant than the ambiguities introduced by the superposition of emission and absorption lines. Similarly, while the assumption of $T_\mathrm{rot} = \SI{2.73}{\kelvin}$ is valid for absorption components arising from diffuse and translucent gas in foreground components, it is likely to be higher for absorption components arising from the envelope of the sources. Therefore, we only report column density upper limits for these components.

\subsubsection{Spectral line contaminants}
In general, spectra observed towards hot cores often show emission from a wide range of molecules, including complex organic species often referred to as \enquote*{weeds} \citep[see, e.g.,][]{Belloche2013}. Features from these species \enquote*{contaminate} the observed profiles, making it difficult to accurately assess the true shape of emission and absorption features, which could lead to significant underestimations of the derived column density values. In the following paragraphs, we briefly discuss the main weeds that contaminate the parts of the spectra relevant to our HCl studies.

In sources with broad outflow components significantly offset from the central velocity, the wings of the \hcl line are blended with the 
SO$^+\,(J = 14\,$--13) line at $\SI{625.9}{\giga\hertz}$. 
Although this blending is observed in only five of the hot-core sources in our sample, we account for the contributions of SO$^+$ by jointly fitting the higher frequency $\Lambda$-doublet component of the same rotational transition at $\SI{626.2}{\giga\hertz}$ \citep{Amano1992,pickett1998jpl}, which has similar spectroscopic parameters and is covered in the same sideband of our observations. 

Similarly, the outflow components of the \hcliso line are blended with the HFS components of CH$_3$CN$\,(J = 34\,$--$ 33)$ and CH$_3$OCH$_3\,(J = 16\,$--$ 15)$, which lie near $\SI{624.9}{\giga\hertz}$. However, their contributions are not modelled but rather masked, as the outflow components of \hcliso are less prominent, causing its red-shifted outflow wing to fall below the noise level in the observed line profiles. Additionally, the \hcliso lines may be contaminated by the SiH$\,(J = 2\,$--$ 1)$ line at $\SI{624.9}{\giga\hertz}$ \citep{schilke2001line, Cernicharo2010}. However, since SiH has never been unambiguously detected in the ISM, any contributions it may have to the HCl spectra are considered negligible.

\subsection{Spectral line profiles}
\label{sec:spectral-line-profiles}
\begin{figure*}
    \centering
    \includegraphics[width=0.4\textwidth]{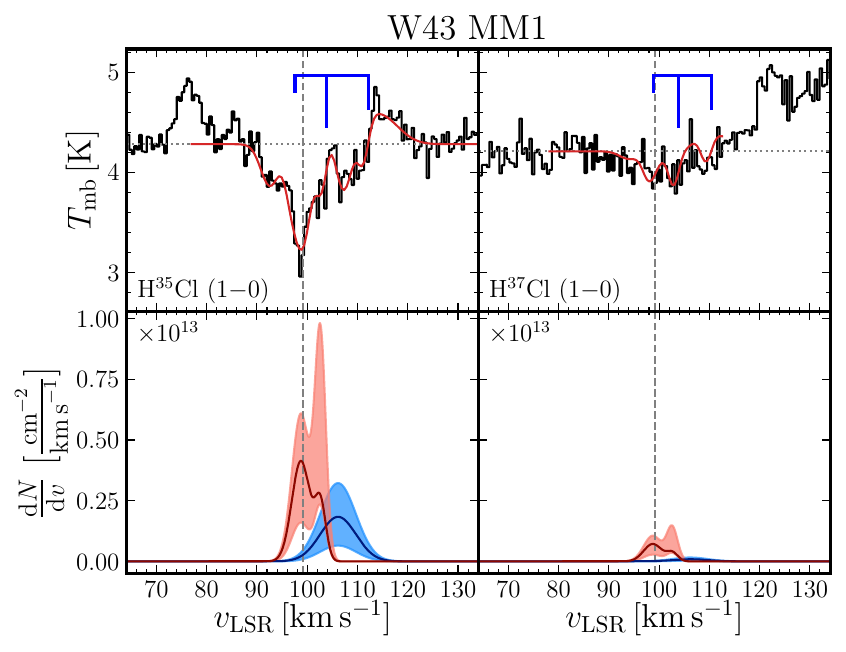}
    \includegraphics[width=0.4\textwidth]{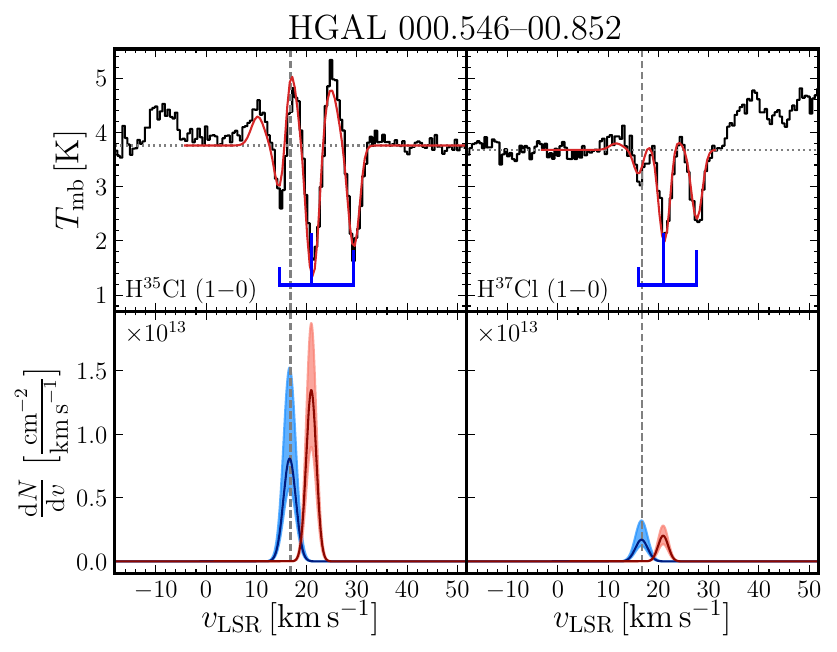} \\
    \includegraphics[width=0.4\textwidth]{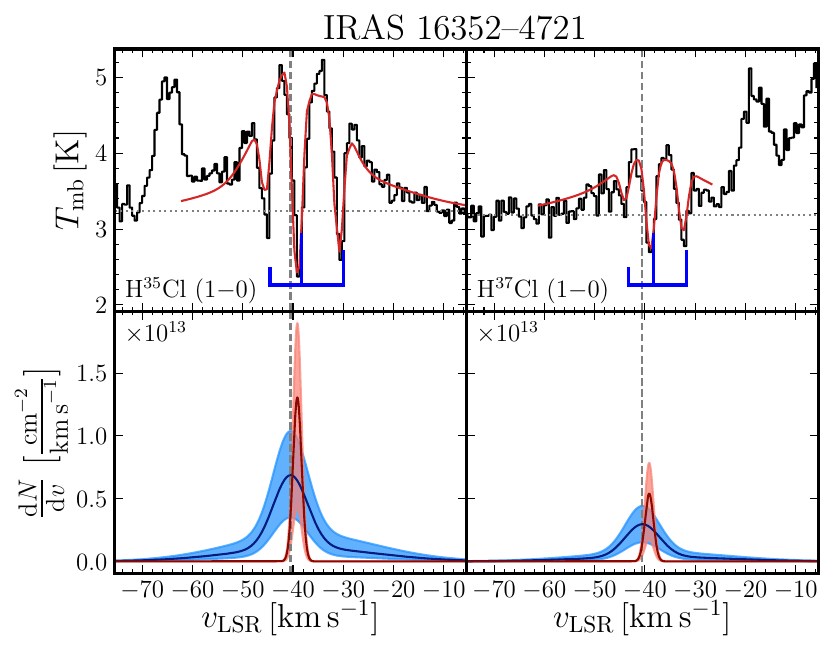}
    \includegraphics[width=0.4\textwidth]{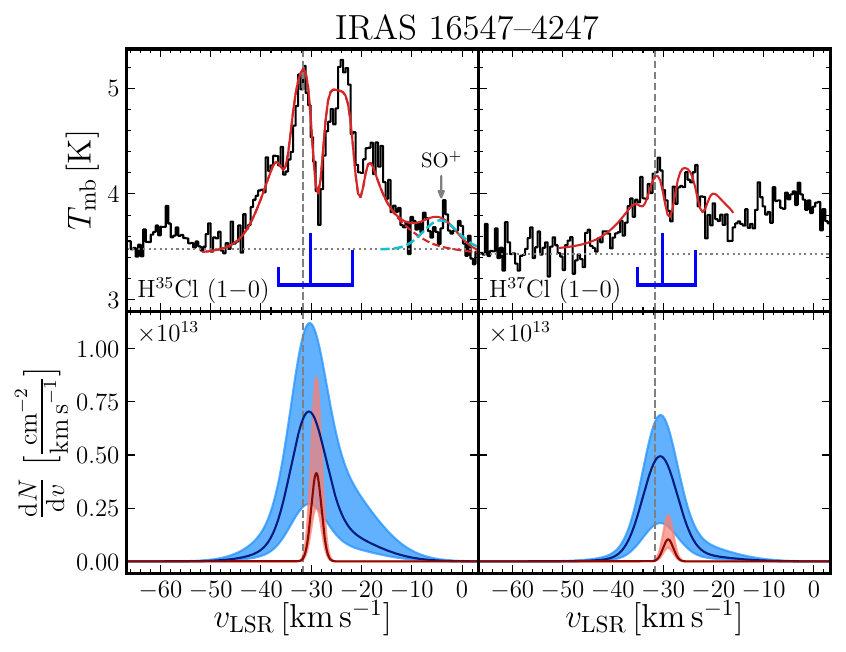}
    \caption[HCl line profiles towards W43-MM1, HGAL~0.546, IRAS~16352 and IRAS~16547]{Upper panel: Observed line profiles of the \hcl and \hcliso\,$J=1$--0 line towards W43-MM1, HGAL~000.546$-$00.852, IRAS~16352$-$4721 and IRAS~16547$-$4247, along with their XCLASS fitted models (red curves). The grey dashed lines mark the systemic velocities, while the grey dotted lines mark the fitted continuum temperatures around the presented lines for each source.
    Bottom panel: The hyperfine structure splitting deconvolved column density profiles. The derived d$N$/d$\upsilon$ distributions with the corresponding 1$\sigma$ errors are displayed by the blue solid curve and blue shaded region, respectively.
    The absorption components are analogously illustrated in red.}
    \label{fig:pcygni-examples}
\end{figure*}

Figures \ref{fig:emission-examples}, \ref{fig:pcygni-examples}, \ref{fig:remaining-emission-fits}  and~\ref{fig:remaining-complex-fits} present all observed spectral line profiles, along with the spectral fits and the resulting column density profiles obtained using XCLASS (where applicable). Tables~\ref{tab:xclass-results-emission} and~\ref{tab:xclass-results-absorption} summarise the line parameters obtained from the spectral fits.

\subsubsection{Emission line profiles}
\label{sec:pure-emission}
\begin{figure*}[htb]
    \centering
    \includegraphics[width=0.4\textwidth]{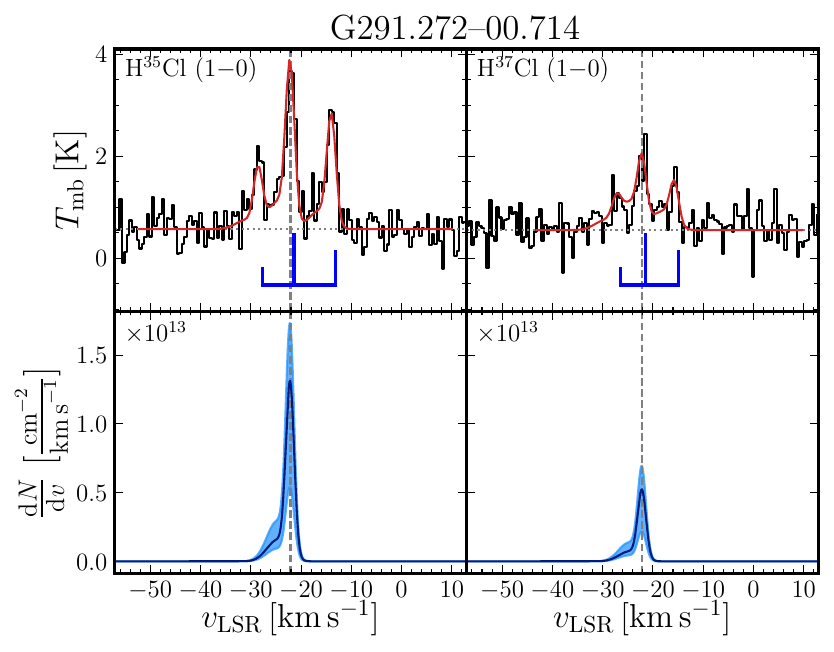}
    \includegraphics[width=0.4\textwidth]{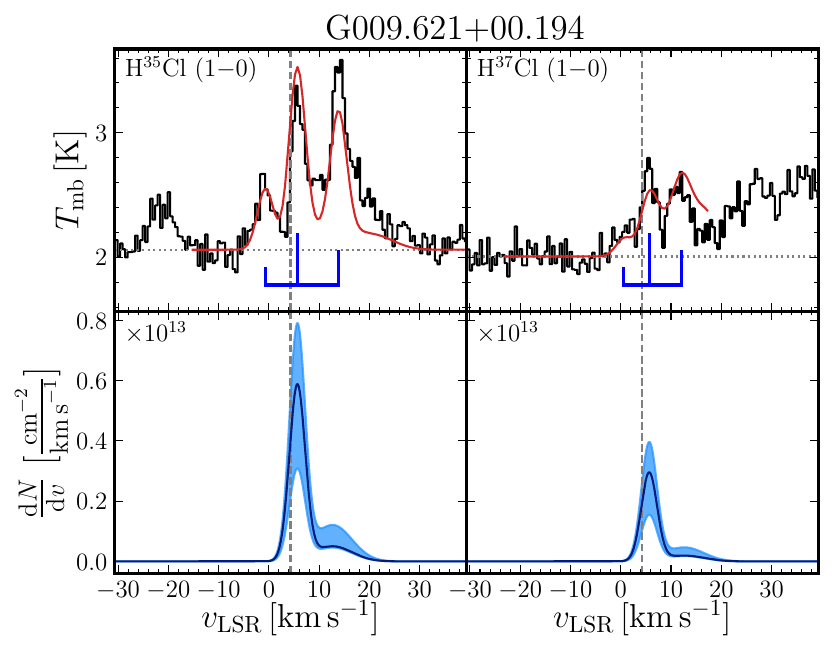} \\
    \includegraphics[width=0.4\textwidth]{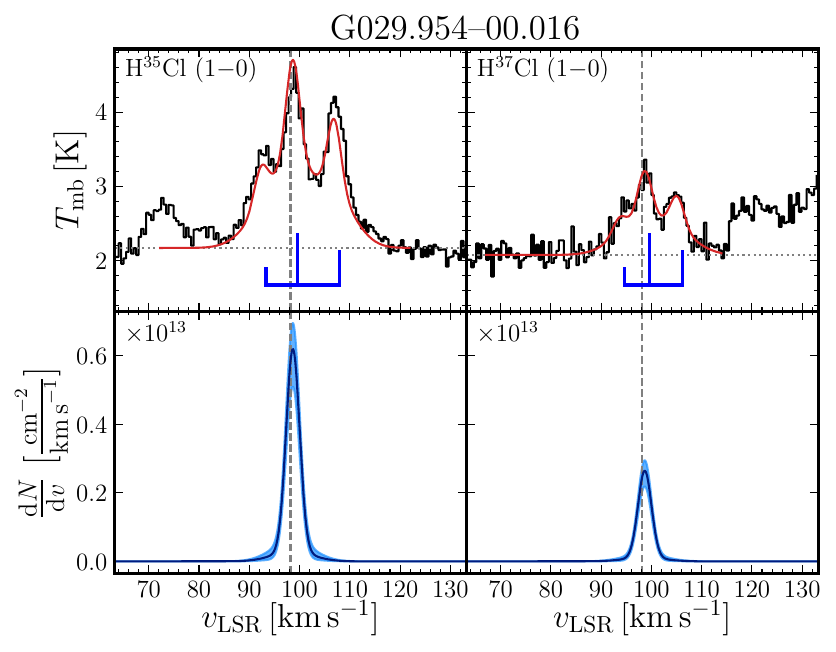}
    \includegraphics[width=0.4\textwidth]{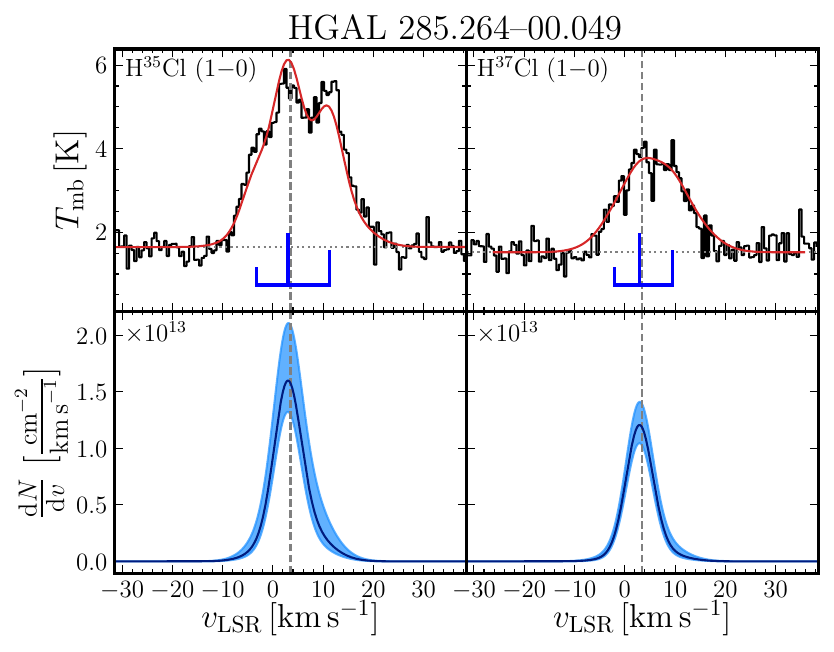}
    \caption[HCl line profiles towards G291.27, G9.62, G29.95 and HGAL~285.26]{Upper panel: Observed spectral line profiles of the \hcl and \hcliso\,$J=1$--0 transitions towards G291.272$-$00.714, G009.621+00.194, G029.954$-$00.016 and HGAL~285.264$-$00.049, along with their XCLASS fitted models (red curves). The grey dashed lines mark the systemic velocities, while the grey dotted lines mark the fitted continuum temperatures around the presented lines for each source.
    Bottom panel: The hyperfine structure splitting deconvolved column density profiles. The derived d$N$/d$\upsilon$ distributions with the corresponding 1$\sigma$ errors are displayed by the blue solid curves and blue shaded regions, respectively.
    }
    \label{fig:emission-examples}
\end{figure*}
The HCl spectra were observed in emission (without absorption components) toward 11 sources among which both isotopologues are detected above a 3$\sigma$ level in all but two targets, G291.579--00.431 and G305.209+00.206.
Lacking a clear detection, the spectral line fitting was not performed on the  spectra toward these sources but their spectra are provided for completeness.

The spectra toward four exemplary targets are displayed in Fig.~\ref{fig:emission-examples}. Assuming only one emission component, we cannot model the observed line profiles as at first glance the line intensity ratio between the HFS components deviate from predictions at LTE, of 1:3:2, (as indicated in the plots in blue). 
Instead, their line shapes indicate that HCl emission originates from multiple components: not only from the dense, inner cores of the background sources but also from the surrounding envelopes or outflowing material. To constrain the multi-component fits, the centroid velocities of the core component are cross-referenced with those of other species, such as the $J=7$--6 line of the optically thin high-density tracer $\mathrm{H^{13}CO^+}$ at $\SI{607.174}{\giga\hertz}$, also covered in our receiver setup. For a subset of our targets, \vLSR could not be constrained using $\mathrm{H^{13}CO^+}$ due to contamination from p-H$_2$O$^+$ absorption. For these sources, systemic velocities, determined from CO isotopologues, from \citet{giannetti2014atlasgal} and \citet{ urquhart2007rms} were used.

In some sources, the observed line intensities remain close to the LTE ratio due to either weak outflow emission (e.g., G291.272$-$00.714) or a broad, symmetric outflow component that uniformly elevates all HFS lines (e.g., G029.954$-$00.016). In these cases, the line profiles are well modelled by the XCLASS fits.
Other sources (e.g., G009.621+00.194), however, show strong deviations from the LTE ratio displaying a `ladder-like' profile with line intensities increasing stepwise resulting in asymmetric profiles with predominately redshifted outflow wings. The XCLASS fitting yielded velocity offsets of $1.5$ to \kms{7.0} and line widths of $8$ to $\sim$\kms{15} for the outflow components. 

While the fits reproduced the general structure of the observed line profiles, some limitations remain, as the relative intensities of the HFS lines arising from the outflow components show minor deviations from LTE.
In the case of HGAL$\,$285.264$-$00.049, for example, the broad outflow wing can be modelled within the noise level, but the absolute intensities of the line features are not entirely reproduced. 
Nevertheless, the fit results provide a consistent, source-by-source description of the observed profiles and their kinematic properties. In conclusion, the emission profiles cannot be modelled satisfactorily without including a broad outflow component, and the derived $\Delta\upsilon$ and $\upsilon_\mathrm{off}$ values reflect the diversity in line kinematics across the sample.
The derived \hcl column densities are generally comparable among the sample, with typical values derived toward the core component on the order of $10^{13}\,\mathrm{cm^{-2}}$. The outflow components exhibit column densities ranging from $\sim$$\SI{10}{\percent}$ to $\SI{70}{\percent}$ of those in the respective cores. No significant optical depth effects are identified in either isotopologue.

\subsubsection{Complex spectral line profiles}
\label{sec:results-absorption}

As discussed above, the HCl line profiles toward the remaining sources are observed in absorption or as a complex combination of both emission and absorption\mychange{, arising due to the complex density and temperature structures of these sources.
To guide the discussion, these spectral line profiles are grouped by their dominant spectral line features into the following categories:
\begin{itemize}
    \item absorption-dominated: Deep absorption against the continuum, often with weak velocity-offset emission components producing clear (inverse) P Cygni shapes (W43-MM1, G351.581$-$00.352, Sgr~B2~(M) and W49~N).
    \item mixed (\enquote*{zigzag}): Alternating emission and absorption with small velocity offsets, yielding a mild P Cygni–like or \enquote*{zigzag} pattern; several cases require an additional outflow component to reproduce the broad underlying shoulder (HGAL$\,$000.546$-$00.852, IRAS~16352$-$4221, G031.412+00.307, G034.258+00.154, G327.293$-$00.579, IRAS~16060$-$5146, G337.704$-$00.054, G010.472+00.027, and G019.609$-$00.234).
    \item emission-dominated (with potential self-absorption): Dominant emission with narrow central absorption dips, indicative of optical depth or self-absorption effects rather than distinct foreground gas.  (IRAS~16547$-$4247, G032.797+00.191, and to a lesser extent W31~C).
\end{itemize}}
The fitting of these profiles is further complicated by the likely presence of multiple emission components, which may arise from both the core and outflow regions (see Sect.~\ref{sec:pure-emission}). Some of these components may remain undetected because of the strength of the absorption, making it difficult to accurately determine the true peaks and depths of the emission and  absorption profiles, respectively. The variety in the observed spectral line profiles arises because of differences in the density structure, temperature, and geometries of these sources. While a quantitative analysis of these differences would require detailed non-LTE radiative transfer analysis, it is presently outside the scope of this work. In the following, we briefly describe the main characteristics of these profile types in more detail.

The absorption-dominated HCl spectra show emission clearly offset in velocity from the absorption dips. This offset is observed relative to the systemic velocities of the sources and results in (inverse) P~Cygni profiles. While multiple distinct absorption components can be identified from the HFS of narrower absorption lines, as seen in the case of G351.581$-$00.352, the profiles of the broader absorption features as seen toward Sgr~B2~(M) and W49~N, do not exclude the possibility of multiple blended components. Similarly, we cannot exclude the presence of emission components arising from several sub-condensations or cores within our beam, which is the case for, e.g., G010.472+00.027 (a source which displays a \enquote*{zigzag} feature, discussed in more detail below), where a satisfactory line fitting could not be achieved. This is particularly evident in the spectra toward W49~N, where the absorption is nested between two emission profiles. Previous studies, such as that by \citet{giannetti2014atlasgal} on CO isotopologues, have revealed the presence of two components at velocities of \kms{4.9} and \kms{12.8}, based on which we model the observed HCl spectra. However, due to uncertainties in the fit and the nature of the emission component, it is challenging to accurately determine both the strength of the emission and the depth of the underlying absorption feature. Overall, while we can fit the observed spectra satisfactorily, additional data are needed to precisely quantify these values.

\mychange{The next set of spectral line profiles corresponds to the mixed (\enquote*{zigzag}) category, arising from emission and absorption features with slight velocity offsets. With the core components typically exhibiting narrow emission, the slight velocity offsets between emission and absorption allow us to reliably model the complex line profiles. Often, an additional emission profile or outflow component is fitted for sources where the \enquote*{shoulder} of a broad underlying emission feature is observed. The clear distinction between the emission and absorption components allows for a more reliable interpretation of the resulting isotopic ratio of the core component. However, even this is not without ambiguities given that there may be additional optical depth effects.}

\mychange{Finally, a few emission-dominated profiles also show narrow central absorption dips, indicative of weak self-absorption, and therefore requiring an additional absorption component in the fit. These cases indicate that, although most emission spectra are well reproduced without absorption, some exhibit potential optical depth effects or self-absorption.}

While the intrinsic line width of the absorption components is typically narrow ($\sim$\kms{3}) for most sources, very broad absorption features are also observed. These broad features, though treated as a single component in the fit due to limited spatial information, may actually result from multiple superimposed absorption components. Although the line fits, based on goodness-of-fit ($\chi^2$) analysis, appear satisfactory, the derived \Ntot values are highly uncertain. These uncertainties primarily stem from difficulties in disentangling the emission and absorption components, as well as assumptions about the rotational temperatures used in modelling both. As a result, the derived \Ntot values likely represent lower limits.

\section{Discussion}
\label{sec:discussion}
In the following section, we derive the chlorine isotopic ratios across a subset of our observed sources. 
We compare the resulting ratios, discuss the impact of optical depth and non-LTE excitation conditions, and consider potential impact of local isotopic fractionation. Our results are then compared to previous HCl observations and predictions from Galactic chemical evolution models.
Finally, we use the kinematic information from the HCl line profile models, together with insights from chemical models, to qualitatively discuss the  the emission of HCl in outflows.

\subsection{Chlorine isotopic ratio across the galaxy}
\label{sec:chlorine-iso}
\begin{figure}[htb]
    \centering
    \includegraphics[width=\columnwidth]{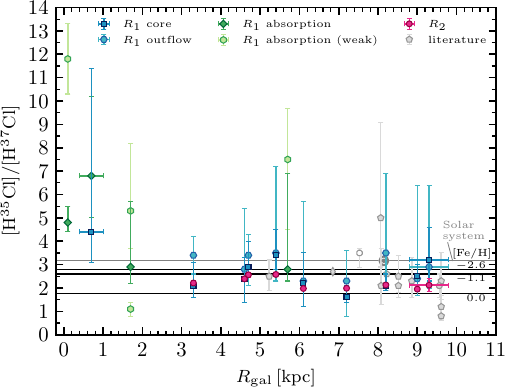}
    \caption[{[H$^{35}$Cl]/[H$^{37}$Cl]} ratio across galactocentric radii]{Chlorine isotopic ratio, [$\mathrm{H^{35}Cl}$]/[$\mathrm{H^{37}Cl}$], as a function of galactocentric radius. The ratio is derived using two methods: $R_1$: derived XCLASS column densities (for both core and outflow components), and $R_2$: ratio of integrated line intensities. $R_1$ results are further divided into emission and absorption components.
    Grey symbols indicate the $^{35}$Cl/$^{37}$Cl ratio from previous studies of a) HCl \citet[][pentagons]{peng2010comprehensive}, \citet[][octagon]{cernicharo201035cl}, \citet[][diamond]{lis2010herschel}, and, b) \htclp, \citet[][white-faced octagon]{neufeld2015herschel}.
    For context, the Solar System value \citep[3.13;][]{lodders20094} and Galactic chemical evolution model predictions for varying metallicities  \citep{kobayashi2011evolution} are shown using horizontal lines.
    }
    \label{fig:iso-ratios}
\end{figure}

The chlorine isotopic ratio is derived from observations of \hcl and \hcliso using two methods, similar to \citet{peng2010comprehensive}. The first, $R_1$, is a more sophisticated approach based on the ratio of total column densities obtained from the XCLASS modelled fits (presented in Sect.~\ref{sec:spec-line-analysis-xclass}) as follows,
$$
R_1 = \frac{ N_\mathrm{tot} \left( \mathrm{H^{35}Cl} \right) }{ N_\mathrm{tot} \left( \mathrm{H^{37}Cl} \right) } \,.
$$
This ratio was determined toward all sources where both HCl lines are detected in emission, as well as those sources where the emission and absorption components are well-separated (Sgr~B2~(M), G351.58$-$00.352, W43-MM1, and HGAL~{000.546$-$00.852}). 
For the emission line profiles, the HCl isotopic ratio are analysed separately in the core and outflow components, to investigate potential fractionation effects in the outflow or envelope regions, where gas temperatures are typically higher. Furthermore, the ratios derived for the outflow components generally have higher uncertainties, likely because of the use of Gaussian line profiles to model, which may not fully capture the complexity of the observed emission. 

The column densities of overlapping core and outflow components derived from the LM algorithm may not represent unique solutions. To obtain a more representative estimate of $R_1$, we adopt the mode of the ratio distribution derived from the MCMC sampling of the full parameter space. The resulting distributions are unimodal and approximately Gaussian, supporting this choice as a statistically meaningful outcome.

The second method, $R_2$, is a more straightforward approach based on the ratio of integrated line intensities. 
Although it does not account for optical depth effects or potential local isotopic fractionation \mychange{between core and outflow} and is therefore less precise than $R_1$, it provides a useful consistency check and additional context for the more detailed analysis. 
\mychange{However, the optical depth, derived from the fitted line profiles, was generally low ($\tau<1$; otherwise the target was analysed in the set of complex spectral line profiles) and isotopic fractionation is unlikely for heavy elements like chlorine ($A\geq35$).} 
For sources showing only emission line profiles, the ratio is calculated using the integrated intensities on the main beam temperature scale over a velocity range of \kms{\pm15} around the main hyperfine component of each isotopologue,
$$
R_2 = \frac{ \textstyle\int^{\upsilon_\mathrm{LSR}+\SI{15}{\kilo\metre\per\second}}_{\upsilon_\mathrm{LSR}-\SI{15}{\kilo\metre\per\second}} T_\mathrm{mb}\left( \mathrm{H^{35}Cl} \right) \mathrm{d}\upsilon }{  \textstyle\int^{\upsilon_\mathrm{LSR}+\SI{15}{\kilo\metre\per\second}}_{\upsilon_\mathrm{LSR}-\SI{15}{\kilo\metre\per\second}} T_\mathrm{mb}\left( \mathrm{H^{37}Cl} \right) \mathrm{d}\upsilon }\,\mathrm{.}
$$
With an end-to-end velocity separation of $\sim$\kms{14}, the integration window over $\pm\SI{15}{\kilo\metre\per\second}$ centred around the main HFS line not only guarantees coverage of all three HFS lines but also outflow features where present. This range, however, slightly excludes the wings of strong ($\gtrsim$\kms{10}) outflow components, but evades contamination from species like, for e.g., SO$^+$.

The errors in $R_1$ are estimated with XCLASS using the MCMC method, as explained in Sect.~\ref{sec:spec-line-analysis-xclass}, while those in $R_2$ are propagated from the error in each channel, that is, the root-mean-square (RMS) noise of the data. 
The HCl isotopic ratios derived using both methods ($R_1$ and $R_2$, as applicable) are listed alongside their XCLASS fit results in Tables~\ref{tab:xclass-results-emission} and~\ref{tab:xclass-results-absorption}. 

Figure~\ref{fig:iso-ratios} presents the results of the $[\mathrm{H^{35}Cl}]/[\mathrm{H^{37}Cl}]$ ratios derived using both methods described above. For comparison, we also included estimates of the chlorine isotopic ratios derived by \citet{peng2010comprehensive}, \citet{cernicharo201035cl}, \citet{lis2010herschel} and \citet{neufeld2015herschel} obtained using HCl except for the latter which used \htclp.

The resulting ratios derived using the first method, $R_1$, show coinciding values for the core and outflow components. This is expected, as the chlorine in both regions likely originates from the same formation site or reservoir, the observed lines are not optically thick, and local isotopic fractionation -- if present -- is minimal for heavy atoms like chlorine ($A\geq35$).
However, the $R_1$ values diverge in several sources with strong outflows, namely HGAL~285.264$-$00.049, IRAS~12326$-$6245 and G009.621+00.194 (located at $R_{\rm gal} = 8.2$, $7.2$ and $\SI{3.3}{kpc}$, respectively). 
In addition to uncertainties from imperfectly modelled line fits, the assumed excitation temperature ($T_\text{ex} = T_\text{dust}$) is likely too low for the hot outflow components, introducing further uncertainties which is reflected in the larger error bars.

In sources exhibiting only emission line profiles, the values of $R_2$ lie between $2.0$ and $2.6$, which fall systematically slightly lower than $R_1$. This difference can be explained by minor optical depth effects, which are accounted for by the XCLASS models. Nevertheless, $R_1$ and $R_2$ remain consistent within the uncertainties.

In sources with absorption components, a larger scatter in $R_1$ is observed. These sources display more complex kinematics, requiring additional components in the modelling. Notably the stronger components (those with higher $N_\text{tot}$), have isotopic ratios that are in agreement with theoretical predictions.
Outliers are observed only among the weaker components, which also carry larger uncertainties.

Higher chlorine isotopic ratios (up to $6.8^{+3.4}_{-1.8}$ for the stronger components) are observed toward the Galactic centre (at $R_\mathrm{gal} < \SI{1}{kpc}$), a trend consistent with that seen in other elements such as carbon and oxygen, and indicative of increased nuclear processing in the inner Galaxy. However, some of the fits presented in this work, are subject to uncertainties arising from low signal-to-noise ratios in some \hcl and \hcliso components, difficulties in disentangling overlapping emission and absorption features, and contributions from multiple cloud components along the line of sight.
Specifically for Sgr~B2~(M), optical depth effects, including the saturation of the HCl absorption line profile, also contribute to the uncertainties in the fit and subsequently derived column densities. 
The overestimation of $R_1$ in the weaker absorption component of W43-MM1 at $\SI{5.7}{kpc}$ is also attributed to similar ambiguities. 

The $^{35}$Cl/$^{37}$Cl isotopic ratio derived toward all sources displaying only emission reproduce values that are in agreement with modelled predictions of chemical yields from supernovae \citep[][as summarised in \citet{maas2018chlorine}]{kobayashi2006NuPhA.777..424N,chieffi2013ApJ...764...21C,kobayashi2011evolution}.
The isotopic ratios of HCl show a scatter with no systematic trend across galactocentric radii, in good agreement with that found by \citet{peng2010comprehensive}, which are determined similar to $R_1$ in this work. Except for two of their targets located at $R_\mathrm{gal}\simeq\SI{9.6}{kpc}$ (with ratios $<1.5$), all targets show, within error bars, isotopic ratios higher than $1.8$, the value predicted from Galactic chemical evolution models with $[\mathrm{Fe/H}] = 0$ from \citet{kobayashi2011evolution}. These authors have included updated yields of AGB stars and CCSNe, the primary formation sites of chlorine, into their models. 
The horizontal lines in Fig.~\ref{fig:iso-ratios} indicate predictions from their models with lower metallicities, $[\mathrm{Fe/H}]<0$, demonstrating that a slight increase in the $^{35}$Cl/$^{37}$Cl isotopic ratio is associated with a lower metallicity, or an earlier stage of Galactic evolution. 
 
The Solar isotopic ratio of chlorine of $3.13$ is higher than the error-weighted average of our sample, $\bar{R}_1 = 2.6\pm0.8$ 
and $\bar{R}_2 = 2.2\pm0.2$. The errors on both values are dominated by statistical variations in the sample. 
This observation does not necessitate that the Solar System originates from a low-metallicity formation site, as \citet{kobayashi2011evolution} predict average Galactic isotopic ratios and local variations are expected.
It would be more appropriate to compare this to the predicted isotopic ratios originating from specific types of supernova models. For example, the model of a CCSN with an initial mass of $\SI{13}{M_\odot}$ and a metallicity of $Z=0.014$, including rotation, by \citet{chieffi2013ApJ...764...21C} yields a comparable isotopic ratio of $^{35}\mathrm{Cl}/^{37}\mathrm{Cl}=3.40$. Their non-rotating models with the same initial mass and metallicity find even higher ratios up to $4.23$.

The elevated isotopic ratio in the Solar System may be inherited from, or influenced by, supernovae enrichment at the site of the Sun's formation, prior to its migration to its current location in the Milky Way \citep{zucker2022localbubble}. 
Similarly, the unique excitation conditions owing to its nuclear activity can potentially explain the higher ratios seen in the vicinity of the Galactic centre.
Based on our current observations, it is unclear whether the regions, toward which higher isotopic ratios are observed, are regions in the vicinity of supernova remnants. This would be an interesting comparison to cross-validate Galactic chemical models.

\subsection{HCl in outflows}
Our observations reveal that HCl emission originates not only from the dense cores but also traces the associated outflows which are often massive. The most spectacular of which are observed toward {G009.621+00.194}, {G045.071+00.132}, IRAS~{16164--5046}, IRAS~{12326--6245}, and {G034.258+00.154}. Some of which have been identified as hosts of explosive outflows; IRAS~{12326--6245}: \citet{zapata2023one}, {G034.258+00.154}: \citet{isaac2025explosive}.
The kinematic profiles of explosive outflows indicate that they are single, extremely energetic events, often assumed to originate from (proto)stellar mergers or the disruption of massive young stellar systems \citep{zapata2023one, bally2011explosive}. 
However, only seven explosive outflows have been reported in literature thus far (based on the kinematics of CO streamers) of which we find HCl toward four \citep[two from this work, {G005.89$-$0.39} and Orion~KL from][respectively]{peng2010comprehensive, schilke2001line}. This raises interesting questions concerning the origins of HCl in these outflows and its potential use as a tracer of these conditions. 

An enhanced abundance of HCl in these regions is not unfathomable as the proposed merger events responsible for the resulting explosive outflows can in the process enrich the environment with fusion products like chlorine.
The increased HCl emission in the outflow could arise from chlorine released by the explosion or from other processes, such as sublimation of HCl ices from grains into the gas phase \citep{kama2015depletion}. Observations at higher angular resolution, like those provided by ALMA, are necessary to probe its spatial distribution and clarify the origins of HCl in outflows.
With an estimated rate of one event every $\sim$90 years in the Milky Way \citep{zapata2023one}, strong HCl outflows may prove to be valuable tracers of explosive outflows, for which the observations toward IRAS~12326$-$6245, G{034.258+00.154}, Orion~KL \citep{schilke2001line}, and G005.89$-$0.39 \citep{peng2010comprehensive} provide promising initial evidence.

\section{Conclusions}
\label{sec:conclusions}
Taking advantage of the high spectral resolution and unique access to the sub-mm wavelength regime provided by the APEX $12\,$m sub-mm telescope, we present here a comprehensive survey of HCl (and \hcliso) toward 28 star-forming regions across the Galaxy.
Among the 28 targets, 11 exhibit spectra with only emission features, two are dominated by absorption, while the remaining 15 show a mixture of both. Since disentangling the components of the latter requires more detailed observations, the analysis presented here focused on those sources toward which we observed HCl in emission. These line profiles could only be satisfactorily modelled by including in addition to the background core component, a broader component likely associated with an outflow. The distinct kinematics of the two HCl components in the observed line profiles reveal that HCl emission traces both dense cores and outflows of the massive star-forming regions studied.

In some of the sources with the most prominent HCl outflows, ALMA CO and SiO observations \citep{zapata2023one, isaac2025explosive} reveal signatures of recent explosive events \citep{bally2011explosive,zapata2023one}. Their nature, however, remains poorly understood. This connection hints at the potential use of HCl as a tracer of explosive processes in massive star-formation. 

Further, the [\hcl\!\!]$/$[\hcliso\!\!] isotopic ratios, determined for 13 sources including four with absorption features in their line profiles, extend the innermost observed Galactocentric range from $5.2\,\mathrm{kpc}$ \citep{peng2010comprehensive} down to $0.1\,\mathrm{kpc}$. The ratios show no significant trend with Galactocentric radius and agree, on average, with those from the survey by \citet{peng2010comprehensive}. Some outliers, particularly the systematically higher ratios toward the Galactic centre, deviate from this flat trend, but are still consistent with predictions from CCSN models with specific initial parameters \citep{maas2018chlorine}. The majority of the observed isotopic ratios agree with Galactic chemical evolution models \citep{kobayashi2011evolution} at metallicities from $\mathrm{[Fe/H]}=-1.1$ to $\mathrm{[Fe/H]}=0$. Overall, this suggests that most Galactic chlorine was synthesized in CCSNe and AGB stars during periods of lower average metallicity in the Galaxy.

The error-weighted average isotopic ratio from the XCLASS fits ($2.6\pm0.8$) agrees with the ratio of the integrated line intensities ($2.2\pm0.2$). This consistency indicates that the derived average is representative of the core components. The error in both cases is dominated by statistical variations within the sample. Hence, the observed scatter probably reflects intrinsic variations from one source to another. It is yet to be determined whether these variations are linked to the proximity to supernova remnants. 
While our study has significantly increased the number of sources in which HCl has been detected --revealing its unique kinematics in both the core and outflow -- it warrants future interferometric observations for a more comprehensive understanding of its nature.

\begin{acknowledgements}
    We sincerely thank the anonymous referee for their careful review and thoughtful suggestions, which greatly improved the clarity and presentation of this work.
    This publication is based on data acquired with the Atacama Pathfinder Experiment (APEX) under the project id M9523C\_109. APEX is operated by the Max-Planck-Institut f\"ur Radioastronomie. 
    We would like to express our gratitude to the APEX staff and science team for their continued assistance in carrying out the observations presented in this work. 
    We are thankful to the developers of GILDAS/CLASS, XCLASS, and of the Python libraries and for making them available as open-source software. In particular, this research has made use of the NumPy, SciPy, Astropy, and Matplotlib packages. We have made use of NASA’s Astrophysics Data System for literature retrieval and citation management. 
   
    We dedicate this work to Karl Martin Menten -- a mentor and colleague to many of the authors of this paper throughout their careers -- who sadly passed away during the drafting of the paper. His accomplishments made much of the scientific work presented here possible, particularly through his key role in the APEX telescope and his pioneering work on molecular masers, which is crucial to the parallax-based distance measurements.
    He provided an exceptionally supportive and motivating scientific environment, shaping the careers of many. His sincere, kind and encouraging nature will be deeply missed.
\end{acknowledgements}

\bibliographystyle{aa}
\bibliography{refs.bib}

\begin{appendix}
    \onecolumn
    \section{Spectral line profiles toward the other sources}
    \label{sec:additional-hcl-lineprofiles}
    \label{sec:appendix-emission}
    \label{sec:complex-absorption}
    This appendix presents the \hcl and \hcliso spectra for the remaining sources, following the examples of the observed HCl spectra shown in Section~\ref{sec:spectral-line-profiles}. 
    
    \begin{figure*}[htbp]
        \centering
        \includegraphics[width=0.32\textwidth]{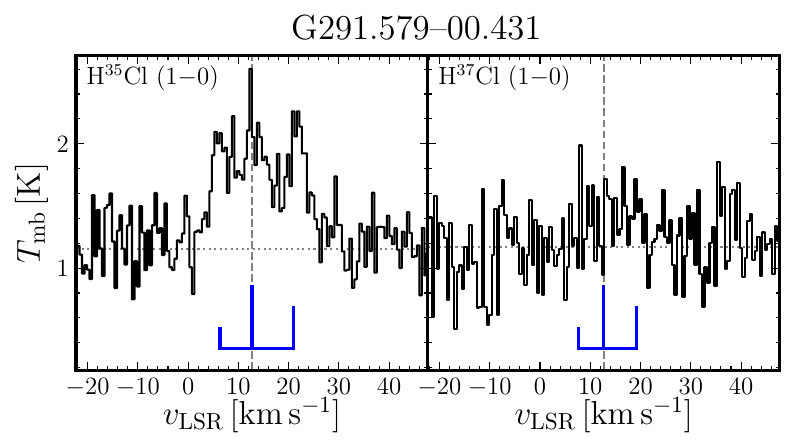}
        \includegraphics[width=0.32\textwidth]{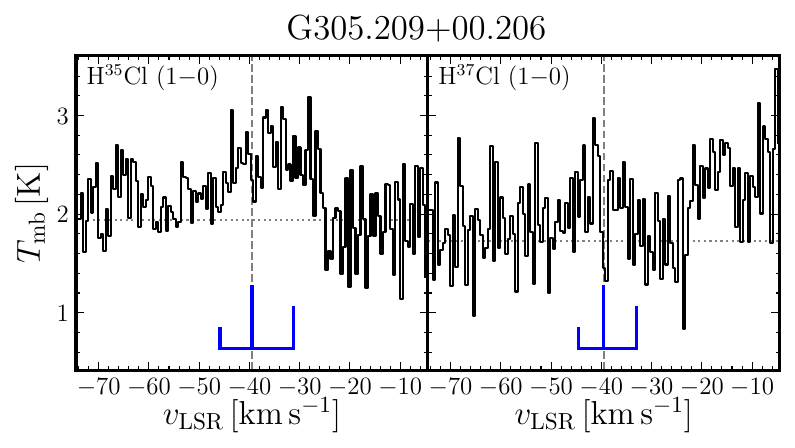}
        \includegraphics[width=0.32\textwidth]{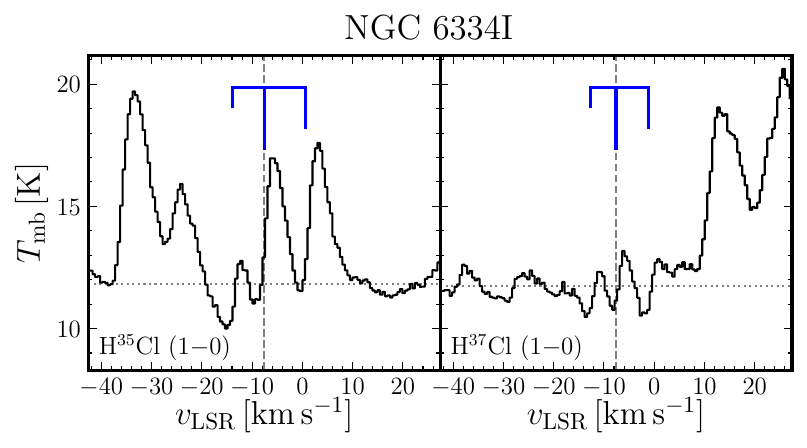} \includegraphics[width=0.32\textwidth]{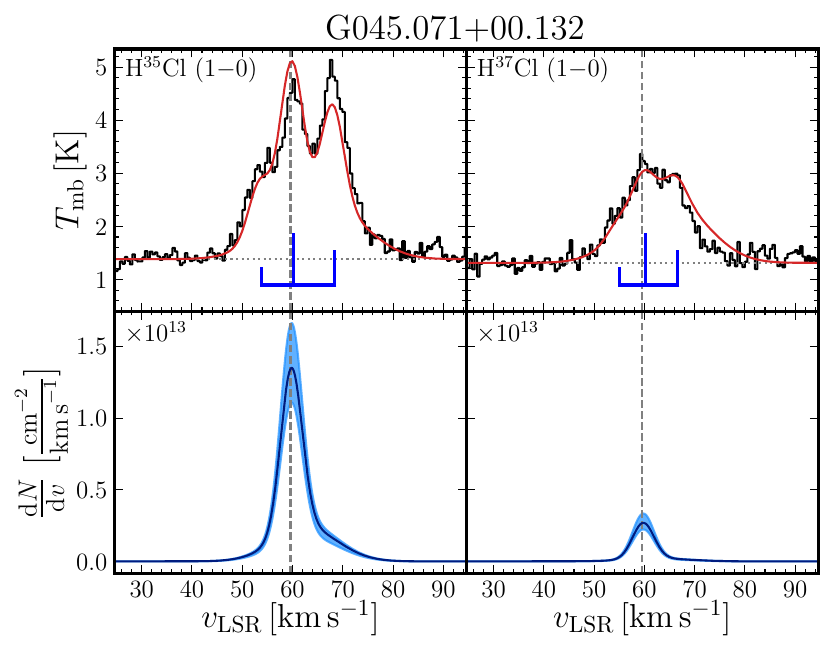}
        \includegraphics[width=0.32\textwidth]{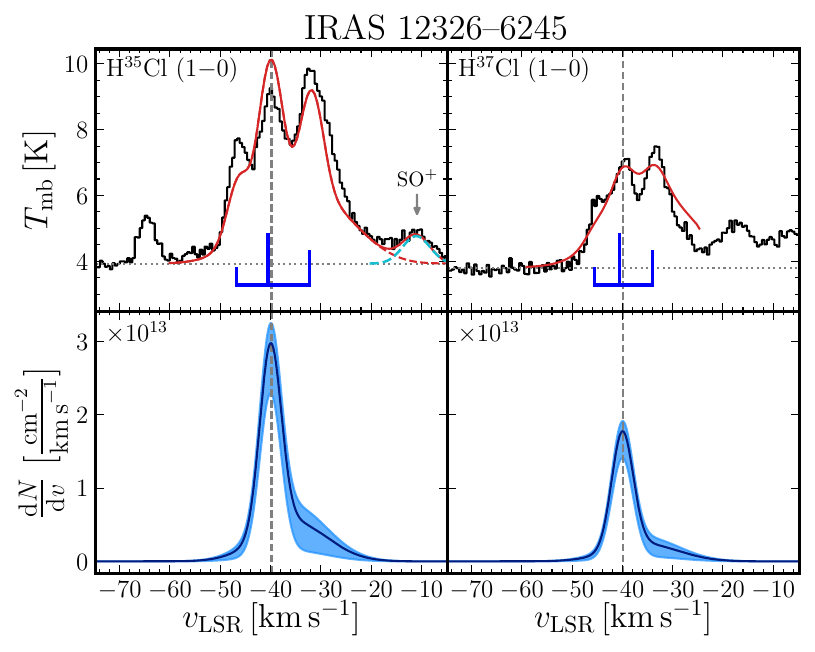}
        \includegraphics[width=0.32\textwidth]{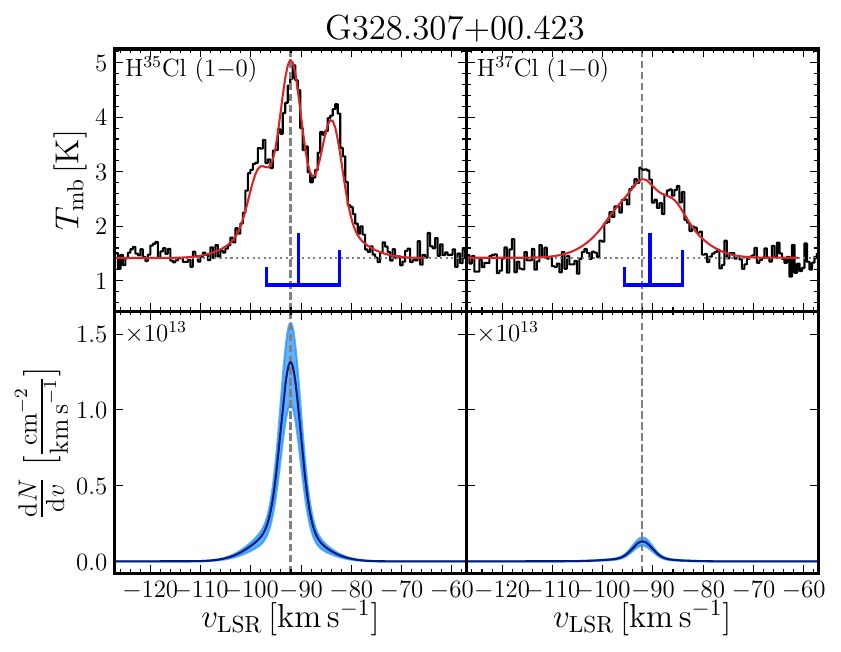}
        \includegraphics[width=0.32\textwidth]{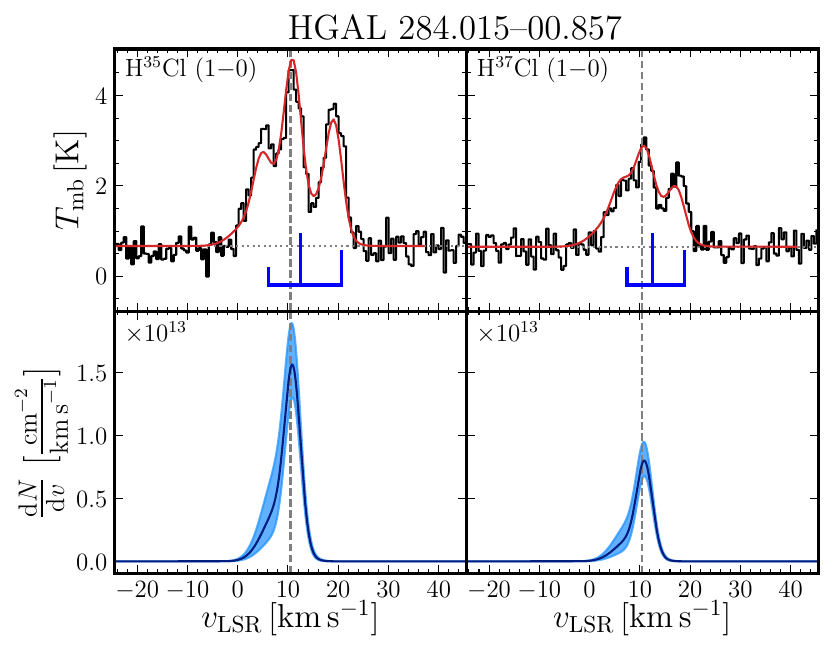}
        \includegraphics[width=0.32\textwidth]{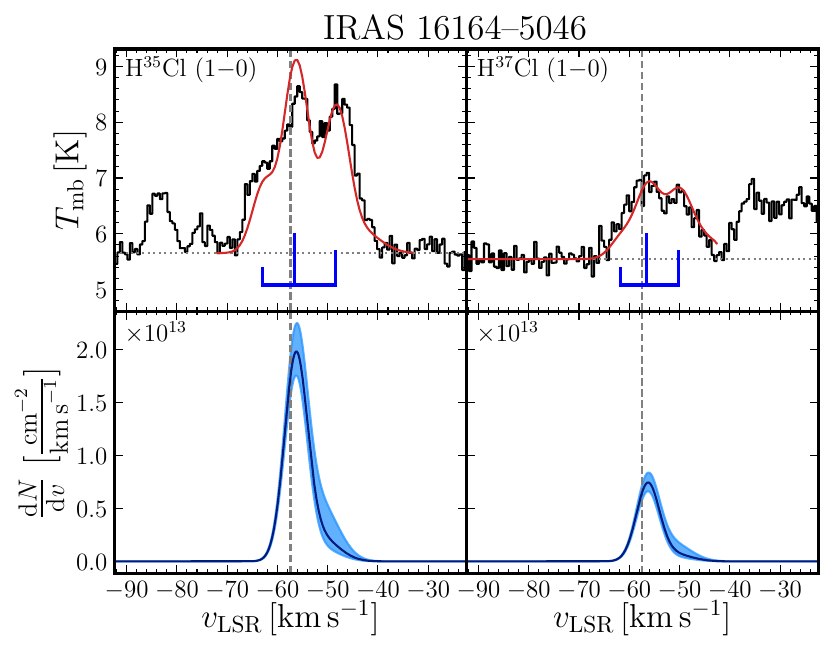}
        
        \caption[HCl emission line profiles]{The observed \hcl and \hcliso\,$J=1$--$\,$0 line spectra toward the remaining sources displaying only emission features, along with a fit (red solid curve), in the upper panels. 
        The hyperfine structure splitting deconvolved column density profiles are illustrated in the lower panel, where d$N$/d$\upsilon$ and the 1$\sigma$ errors for the fitted emission components are displayed by the blue solid curve and blue shaded region. 
        The \hcl and \hcliso spectra, possessing a strong outflow, are contaminated by SO$^+$ and  by CH$_3$CN and CH$_3$OCH$_3$, respectively, towards the \mychange{red} wing. The fits to the SO$^+$ line are shown by the dashed cyan curve. The fit of the contaminating line is overlaid as a dashed cyan curve. The fit to the \hcl spectrum after subtracting contributions from SO$^+$ is displayed by the dashed red curve. The solid red line represents the combined XCLASS fit to both, \hcl and SO$^+$.
        No fit has been performed, if the line temperature of a transition is lower than 3$\sigma$ for the given RMS, or if no baseline could be estimated (NGC\,6334I).}
        \label{fig:remaining-emission-fits}
    \end{figure*}

    \begin{figure*}[htbp]
        \centering
        \includegraphics[width=0.32\textwidth]{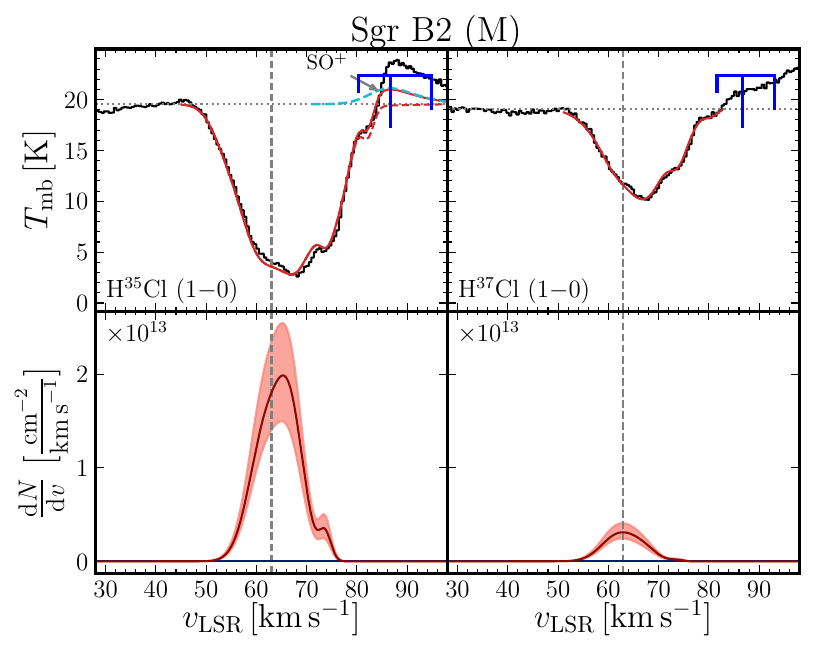}
        \includegraphics[width=0.32\textwidth]{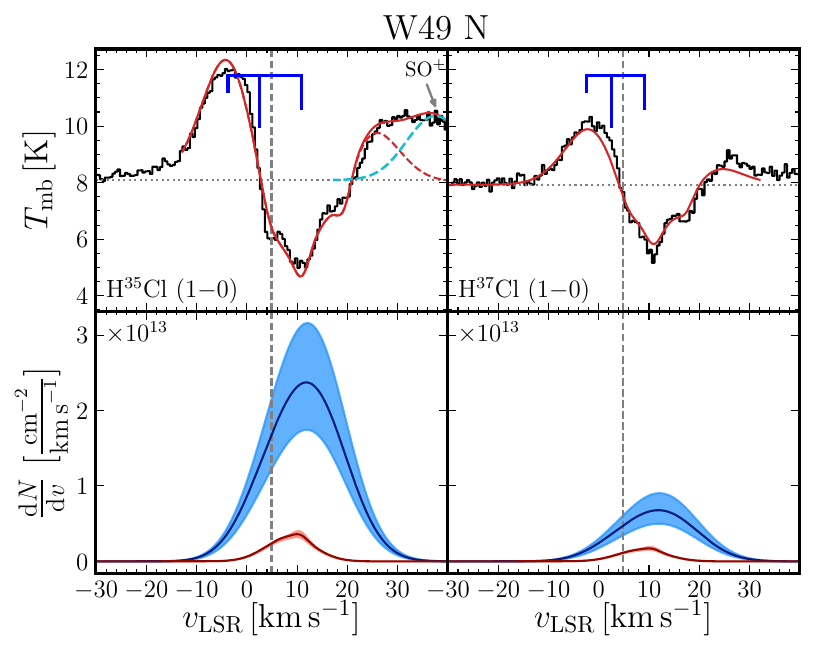} 
        \includegraphics[width=0.32\textwidth]{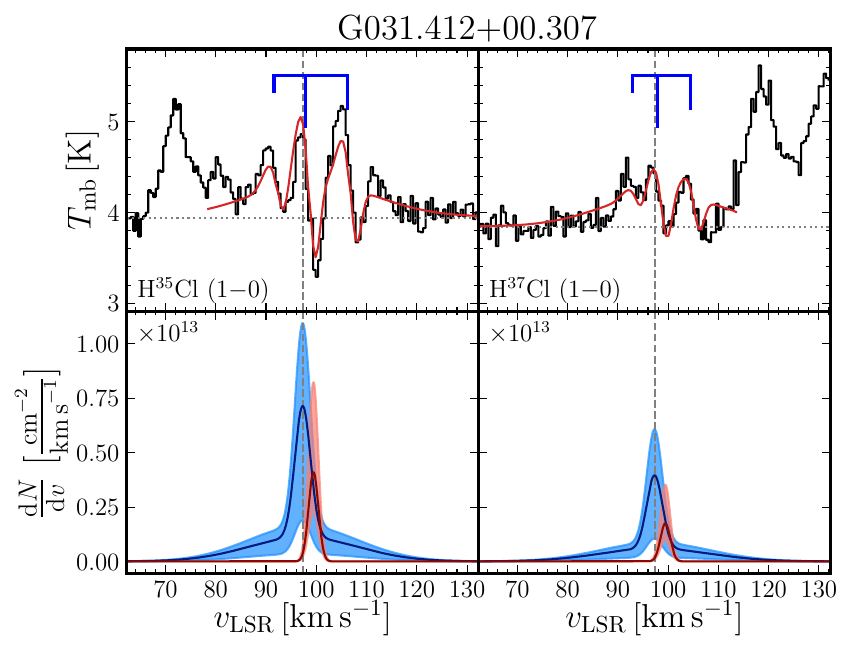} 
        \includegraphics[width=0.32\textwidth]{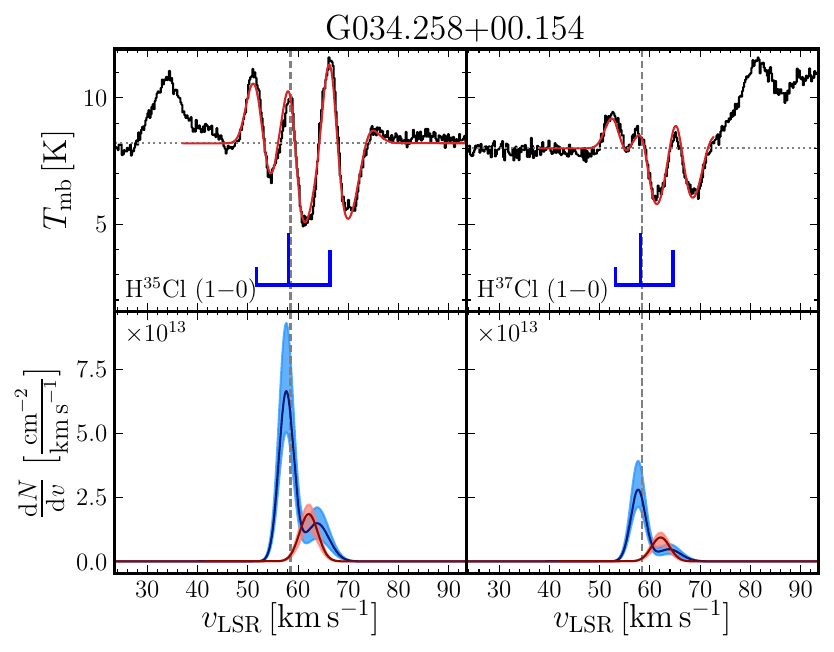} 
        \includegraphics[width=0.32\textwidth]{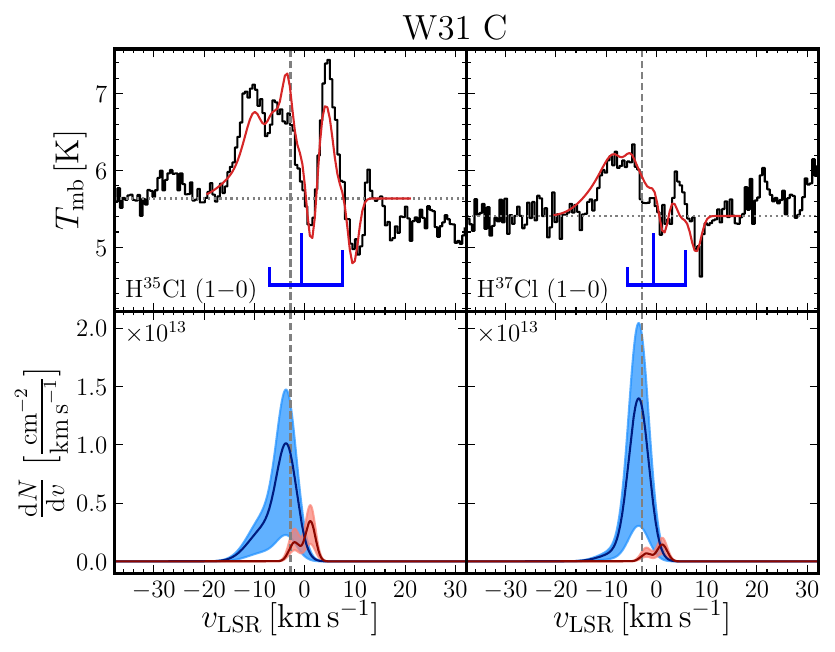} 
        \includegraphics[width=0.32\textwidth]{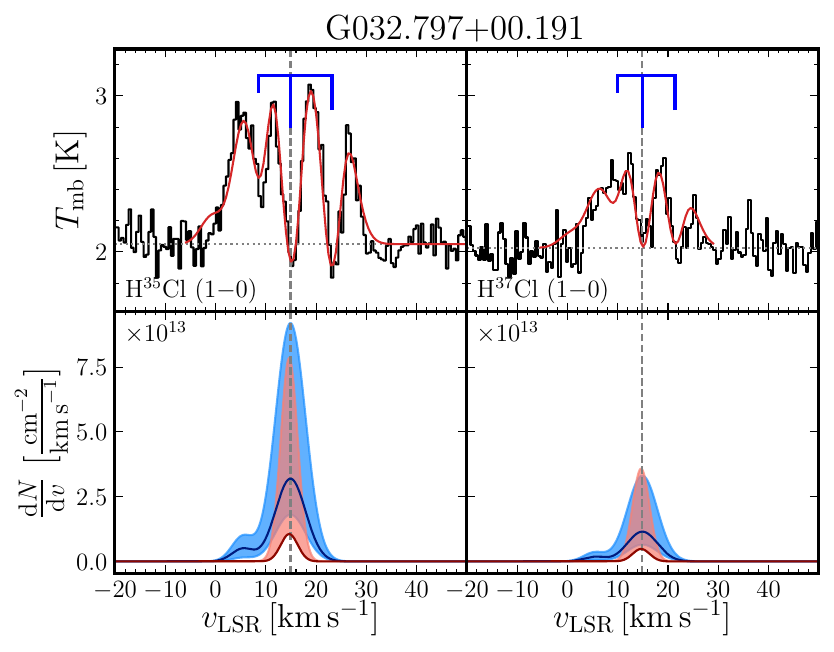}
        \includegraphics[width=0.32\textwidth]{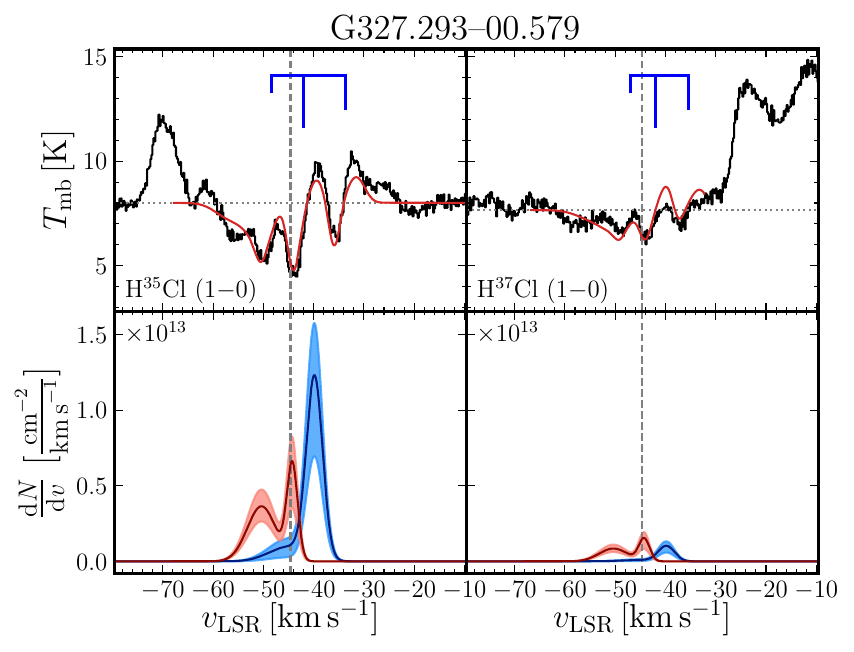} 
        \includegraphics[width=0.32\textwidth]{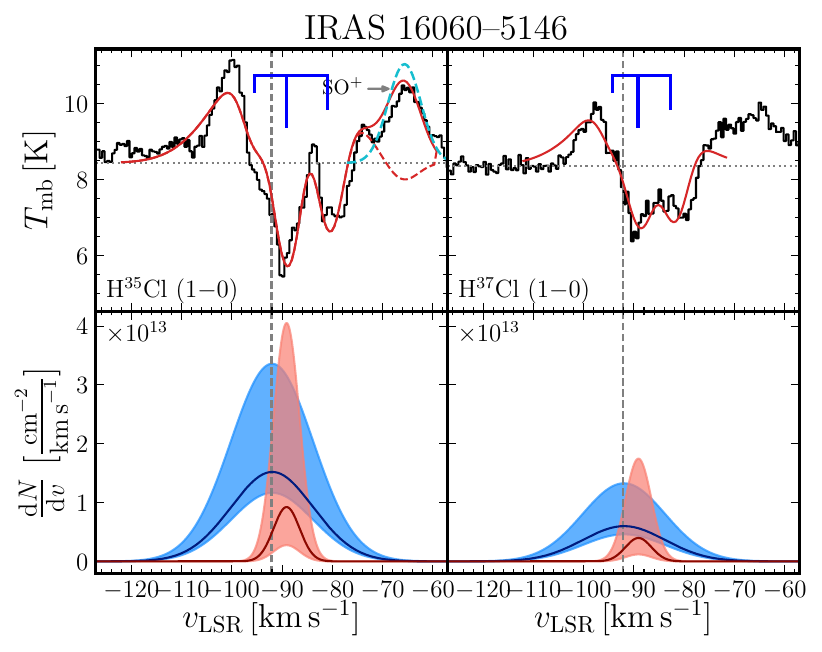}
        \includegraphics[width=0.32\textwidth]{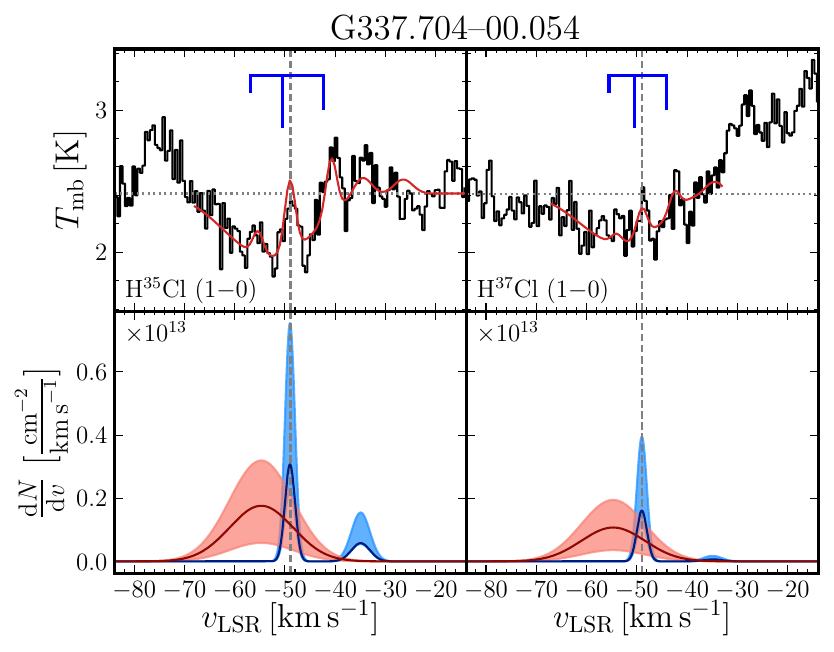}
        \includegraphics[width=0.32\textwidth]{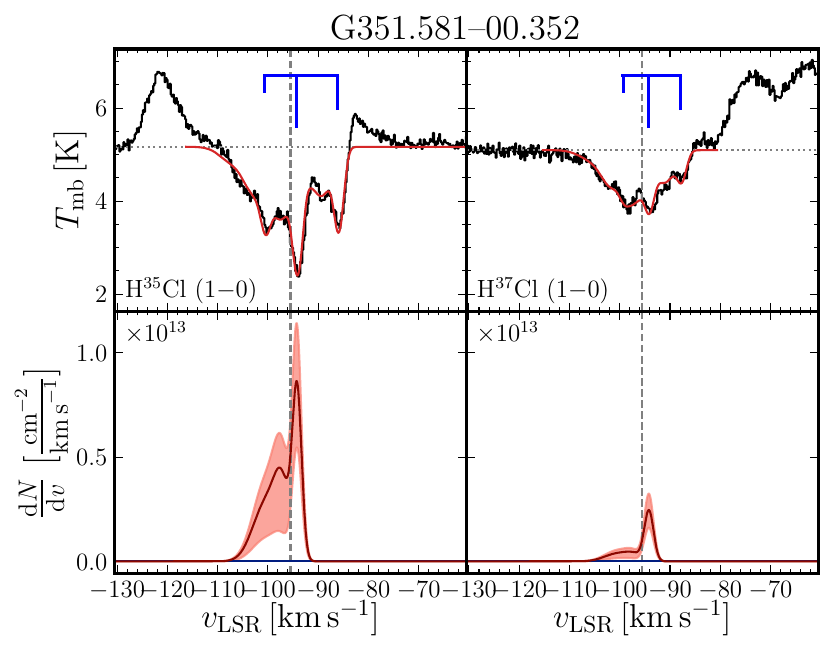} 
        \includegraphics[width=0.32\textwidth]{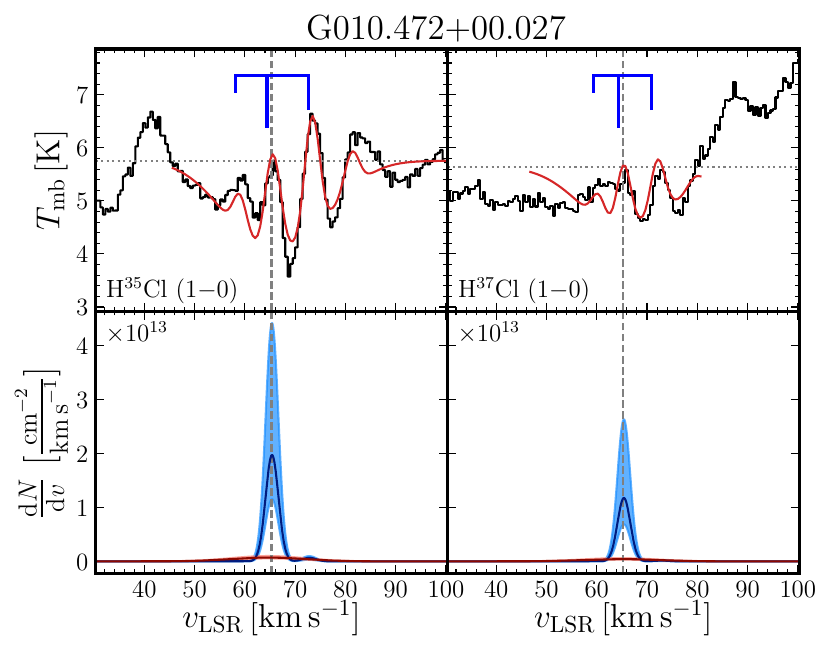}  
        \includegraphics[width=0.32\textwidth]{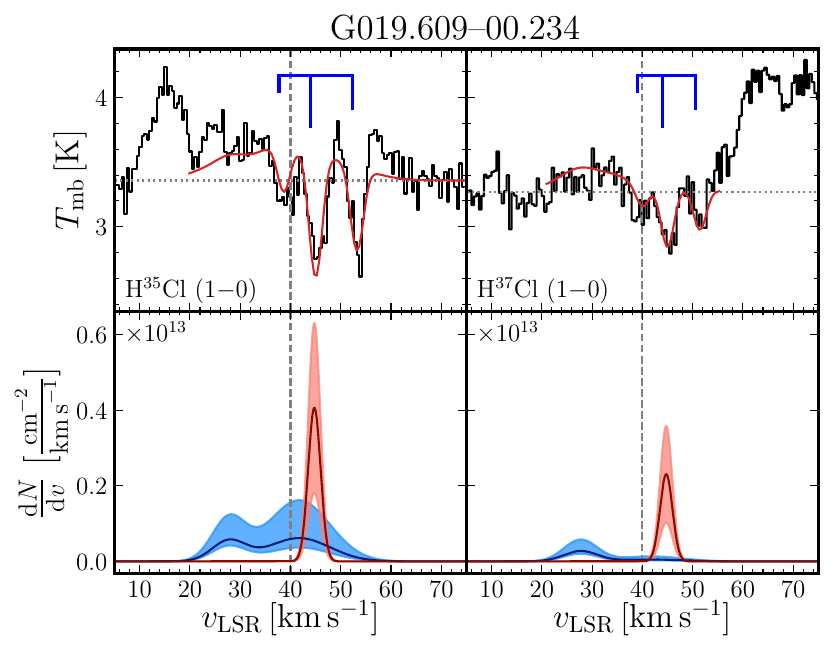}
        \caption[HCl absorption line profiles]{The observed \hcl and \hcliso\,$J=1$--$\,$0 spectra toward the remaining sources displaying complex line profiles, along with a fit (red solid line), in the upper panels. 
        The hyperfine structure splitting deconvolved column density profiles are illustrated in the lower panel, where d$N$/d$\upsilon$ and the 1$\sigma$ errors for the fitted emission components are displayed by the blue solid curve and blue shaded region. 
        The \hcl and \hcliso spectra, possessing a strong outflow, are contaminated by SO$^+$ and  by CH$_3$CN and CH$_3$OCH$_3$, respectively, towards the red wing. The fits to the SO$^+$ line are shown by the dashed cyan curve. The fit of the contaminating line is overlaid as a dashed cyan curve. The fit to the \hcl spectrum after subtracting contributions from SO$^+$ is displayed by the dashed red curve. The solid red line represents the combined XCLASS fit to both, \hcl and SO$^+$.
        No fit has been performed, if the line temperature of a transition is lower than 3$\sigma$ for the given RMS.}
        \label{fig:remaining-complex-fits}
    \end{figure*}
    
    \FloatBarrier
    \onecolumn
    \section{Fitted line parameters from XCLASS}
    \label{sec:xclass-fit-tables}
    The fit parameters resulting from the XCLASS fits, as described in Sec.~\ref{sec:spec-line-analysis-xclass}, together with the errors on them, estimated by the MCMC runs, are presented in Table~\ref{tab:xclass-results-emission} for the targets with pure emission line profiles and in Table~\ref{tab:xclass-results-absorption} for those with absorption line profiles.
    \begin{table*}[htbp]
        \caption{Fitted line parameters from XCLASS for the emission spectra.}
        \label{tab:xclass-results-emission}
        \centering
        \begin{tabular}{ll|cccrcc}
                
            \hline
            \hline
            \multicolumn{1}{l}{ Source} & \multicolumn{1}{l|}{ Component} & \multicolumn{1}{c}{$T_\mathrm{rot}$$^{(a)}$} & \multicolumn{1}{c}{$N_\mathrm{tot}$} & \multicolumn{1}{c}{$\Delta\upsilon$} & \multicolumn{1}{c}{$\upsilon_\mathrm{LSR}$} & \multicolumn{1}{c}{$R_1$$^{(b)}$} & \multicolumn{1}{c}{$R_2$$^{(c)}$} \\ 
            
                                      &            &     [K] & 
            [$\times10^{12}\,\mathrm{cm^{-2}}$] & [$\mathrm{km\,s^{-1}}$] & \multicolumn{1}{c}{[$\mathrm{km\,s^{-1}}$]} && \\
            \hline                          
                  HGAL 284.015–00.857 &      core & $35.0^*$\negmedspace\! &   \hphantom{0}$50.2^{+4.4}_{-4.9\hphantom{1}}$ &   \hphantom{1}$3.6^{+0.4}_{-0.4\hphantom{1}}$ &  $11.0^{+0.2}_{-0.3}$ &  $2.5^{+0.5}_{-0.4}$ & $2.0\pm0.1$ \\
                                      &   outflow & $35.0^*$\negmedspace\! & \hphantom{0}$31.2^{+24.4}_{-15.8}$ &   \hphantom{1}$7.2^{+9.1\hphantom{1}}_{-3.0}$ &   $2.5^{+4.0}_{-0.7}$ &  $2.4^{+4.0}_{-0.7}$ &             \\
                  HGAL 285.264–00.049 &      core & $38.4$ &   \hphantom{0}$79.8^{+4.9}_{-8.4\hphantom{1}}$ &   \hphantom{1}$6.1^{+1.1\hphantom{1}}_{-1.7}$ &   $2.9^{+0.5}_{-0.3}$ &  $2.1^{+0.5}_{-0.2}$ & $2.1\pm0.1$ \\
                                      &   outflow & $38.4$ & \hphantom{0}$50.4^{+57.9}_{-20.1}$ & $12.1^{+10.6}_{-3.5}$ &  $4.4^{+8.2}_{-10.5\negmedspace\negmedspace\!}$ &  $3.5^{+3.4}_{-0.9}$ &             \\
                      G291.272–00.714 &      core & $38.6$ &  \hphantom{0}$25.9^{+7.5}_{-15.5}$ &   \hphantom{1}$2.0^{+0.8\hphantom{1}}_{-0.3}$ & $-22.2^{+4.2}_{-0.3}$ &  $3.2^{+1.4}_{-0.9}$ & $2.1\pm0.3$ \\
                                      &   outflow & $38.6$ &    \hphantom{00}$7.9^{+7.0}_{-3.2\hphantom{1}}$ &   \hphantom{1}$4.8^{+10.3}_{-1.8}$ & $-24.8^{+8.6}_{-7.0}$ &  $2.9^{+3.5}_{-0.9}$ &             \\
                      G291.579–00.431 &      core & $27.1$ &    \mychange{\hphantom{0\!}$<\!2.6^{(d)}_{\hphantom{-5.21}}$}  &                   --- &                   --- &                  --- & $4.7\pm1.3$ \\
                                      &   outflow & $27.1$ &    \mychange{\hphantom{0\!}$<\!3.7^{(d)}_{\hphantom{-5.21}}$} &                   --- &                   --- &                  --- &             \\
                      IRAS 12326–6245 &      core & $34.2$ & $135.0^{+5.3}_{-19.3}$ &    \hphantom{1}$5.0^{+1.1\hphantom{1}}_{-3.2}$ & $-40.0^{+0.4}_{-0.2}$ &  $1.6^{+0.5}_{-0.2}$ & $2.0\pm0.1$ \\
                                      &   outflow & $34.2$ & \hphantom{0}$92.6^{+36.4}_{-68.0}$ &  $16.0^{+5.5\hphantom{1}}_{-5.5}$ & $-35.6^{+1.0}_{-0.2}$ &  $2.3^{+1.3}_{-1.5}$ &             \\
                      G305.209+00.206 &      core & $28.2$ &    \mychange{\hphantom{0\!}$<\!3.9^{(d)}_{\hphantom{-5.21}}$} &                   --- &                   --- &                  --- & $2.1\pm0.5$ \\
                                      &   outflow & $28.2$ &    \mychange{\hphantom{0\!}$<\!5.5^{(d)}_{\hphantom{-5.21}}$} &                   --- &                   --- &                  --- &             \\
                      G328.307+00.423 &      core & $35.7$ & \hphantom{0}$52.6^{+10.0}_{-10.1}$ &   \hphantom{1}$4.6^{+0.8\hphantom{1}}_{-0.8}$ & $-92.1^{+0.4}_{-0.3}$ &  $2.4^{+0.9}_{-1.0}$ & $2.4\pm0.1$ \\
                                      &   outflow & $35.7$ &  \hphantom{0}$32.9^{+7.4}_{-10.3}$ &  $13.1^{+3.9\hphantom{1}}_{-2.1}$ & $-92.8^{+1.5}_{-0.6}$ & $2.8^{+2.6}_{-1.4}$ &             \\
                      IRAS 16164–5046 &      core & $31.4$ & $108.0^{+7.8}_{-10.7}$ &   \hphantom{1}$5.4^{+2.2\hphantom{1}}_{-0.6}$ & $-56.3^{+0.6}_{-0.5}$ &  $3.4^{+1.1}_{-0.6}$ & $2.6\pm0.1$ \\
                      (G332.83)&   outflow & $31.4$ &  \hphantom{0}$19.7^{+28.2}_{-8.8\hphantom{1}}$ &  \hphantom{1}$8.5^{+11.2}_{-2.8}$ & $-51.4^{+9.6}_{-7.0}$ &  $3.5^{+3.7}_{-1.2}$ &             \\
                      G009.621+00.194 &      core & $32.1$ &  \hphantom{0}$22.8^{+7.4}_{-11.0}$ &   \hphantom{1}$3.7^{+9.8\hphantom{1}}_{-1.1}$ &   $5.7^{+4.9}_{-0.5}$ &  $2.1^{+1.0}_{-0.5}$ & $2.2\pm0.1$ \\
                                      &   outflow & $32.1$ &    \hphantom{00}$4.9^{+6.8}_{-0.5\hphantom{1}}$ &  \hphantom{1}$9.1^{+10.9}_{-2.7}$ &  $12.7^{+9.6}_{-4.4}$ &  $3.4^{+0.8}_{-0.8}$ &             \\
                      G029.954–00.016 &      core & $35.5$ &   \hphantom{0}$19.9^{+1.5}_{-3.1\hphantom{1}}$ &   \hphantom{1}$3.2^{+0.9\hphantom{1}}_{-0.5}$ &  $98.7^{+0.3}_{-0.4}$ &  $2.9^{+1.1}_{-0.4}$ & $2.6\pm0.1$ \\
                                      &   outflow & $35.5$ &    \hphantom{00}$2.8^{+3.2}_{-1.9\hphantom{1}}$ &   \hphantom{1}$9.5^{+5.2\hphantom{1}}_{-1.7}$ &  $98.6^{+1.6}_{-0.1}$ &  $3.4^{+0.9}_{-1.3}$ &             \\
                      G045.071+00.132 &      core & $37.0$ & \hphantom{0}$61.0^{+15.2}_{-10.1}$ &   \hphantom{1}$5.0^{+1.7\hphantom{1}}_{-0.6}$ &  $59.8^{+0.4}_{-0.1}$ &  $2.2^{+1.3}_{-1.0}$ & $2.0\pm0.1$ \\
                                      &   outflow & $37.0$ &   \hphantom{0}$36.6^{+5.0}_{-9.4\hphantom{1}}$ &  $14.7^{+4.1\hphantom{1}}_{-2.2}$ &  $62.7^{+1.8}_{-1.0}$ & $2.3^{+3.4}_{-1.1}$ &             \\
            
            \hline
        \end{tabular}

        \tablefoot{ For sources known under multiple designations, the alternative designation not used in this work is provided in italics below the source name. The errors correspond to 1$\sigma$ confidence intervals. The dust temperature of HGAL\,284.015$-$00.857 is not given by \citet{schuller2009atlasgal}, hence it is fixed to $\SI{35}{\kelvin}$, a typical temperature of sources in the sample. $^{(a)}$ XCLASS assumes LTE, such that $T_\mathrm{rot}=T_\mathrm{ex}=T_\mathrm{kin}$, further we assume $T_\mathrm{rot}=T_\mathrm{dust}$ from \citet{schuller2009atlasgal}, $^{(b)}$ $R_1 = \frac{N(\mathrm{H^{35}Cl})}{N(\mathrm{H^{37}Cl})}$, $^{(c)}$ $R_2 = \frac{\int T_\mathrm{mb}\mathrm(\mathrm{H^{35}Cl}){\mathrm{d}}\upsilon}{\int T_\mathrm{mb}(\mathrm{H^{37}Cl}){\mathrm{d}}\upsilon}$ integrated over $\pm\SI{15}{\kilo\metre\per\second}$ around the main HFS line, guaranteeing to cover all three HFS lines. This range, however, slightly excludes the wings of the outflow components, but evades contamination from species like, e.g., SO$^+$. \mychange{$^{(d)}$ 3$\sigma$ upper limits derived for non-detections, assuming optically thin, beam-filled emission and line widths of \kms{5} for the core and \kms{10} for the outflow.}}
        
    \end{table*}
    \begin{table*}
        \caption{Fitted line parameters from XCLASS for the P Cygni and absorption line profiles.}
        \label{tab:xclass-results-absorption}
        \centering
        \begin{tabular}{ll|cccrc}
            \hline
            \hline
            \multicolumn{1}{l}{Source}     &     \multicolumn{1}{l|}{Component}     &     \multicolumn{1}{c}{$T_\mathrm{rot}$$^{(a)}$}     &     \multicolumn{1}{c}{$N_\mathrm{tot}$}     &     \multicolumn{1}{c}{$\Delta\upsilon$}     &     \multicolumn{1}{c}{$\upsilon_\mathrm{LSR}$}     &     \multicolumn{1}{c}{$R_1$$^{(b)}$} \\ 
                                          &                    &         [K]     &     [$\times10^{12}\,\mathrm{cm^{-2}}$]     &     [$\mathrm{km\,s^{-1}}$]     &     \multicolumn{1}{c}{[$\mathrm{km\,s^{-1}}$]}     &     \\
            \hline 
                      G327.293–00.579     &          core     &     $27.5$     &           \,\hphantom{00}$9.2^{+5.0}_{-6.7}$  \hphantom{$^{44}$}   &      \hphantom{1}\!  $9.0^{+2.1}_{-1.4}$     &       $-45.1^{+2.1}_{-3.7}$     &     $1.9^{+1.0}_{-0.9}$ \\
                           &       outflow     &     $27.5$     &       \hphantom{0} $48.0^{+13.0}_{-20.5}$ \hphantom{$^{4}$}    &      \hphantom{1}\!  $3.8^{+0.7}_{-0.6}$     &       $-39.8^{+3.5}_{-0.9}$     &     $2.0^{+1.0}_{-0.9}$                 \\
                           &       absorption     &     $2.73$     &        \hphantom{0}  $25.8^{+7.8}_{-7.1}$  \hphantom{$^{44}$}   &       \hphantom{1}\! $6.7^{+3.1}_{-1.0}$     &       $-50.4^{+1.5}_{-5.0}$     &      $2.0^{+1.5}_{-1.0}$           \\
                           &       absorption     &     $2.73$     &        \hphantom{0}  $16.8^{+4.1}_{-8.0}$  \hphantom{$^{44}$}   &       \hphantom{1}\! $2.5^{+0.3}_{-0.2}$     &       $-44.2^{+3.9}_{-0.4}$     &      $2.0^{+1.5}_{-1.0}$                \\
            IRAS 16060–5146      &          core     &     $32.2$     &      $311.0^{+380.0}_{-72.7}$     &     $19.4^{+12.9}_{-1.4}$\negmedspace\negmedspace     &     $-92.0^{+10.4}_{-10.5}$     &      $2.4^{+1.2}_{-0.9}$  \\
            (G330.95)     &       outflow     &     $32.2$     &       \hphantom{00}    $1.4^{+0.0}_{-0.4}$  \hphantom{$^{44}$}   &       \hphantom{1}\! $8.1^{+5.6}_{-1.0}$     &       $-91.2^{+8.9}_{-9.1}$     &      $2.1^{+1.2}_{-0.6}$                 \\
                         &       absorption     &     $2.73$     &      \hphantom{\,} $60.0^{+203.0}_{-42.1}$     &       \hphantom{1}\! $6.1^{+2.6}_{-2.1}$     &       $-89.1^{+7.1}_{-5.4}$     &      $3.0^{+0.7}_{-1.5}$                 \\
                      G337.704–00.054     &          core     &     $23.6$     &        \hphantom{00}  $7.7^{+11.2}_{-2.0}$   \hphantom{$^{4}$}  &       $\hphantom{1}\! 2.4^{+1.7}_{-0.6}$     &       $-49.0^{+6.4}_{-8.0}$     &      $1.5^{+2.6}_{-0.6}$  \\
                           &       outflow     &     $23.6$     &          \hphantom{00} $2.7^{+4.5}_{-0.1}$   \hphantom{$^{44}$}  &      \hphantom{1}\!  $4.4^{+3.7}_{-1.6}$     &       $-34.9^{+2.7}_{-3.5}$     &      $10.7^{+0.7}_{-3.8}$                 \\
                           &       absorption     &     $2.73$     &       \hphantom{0} $28.0^{+22.8}_{-18.7}$   \hphantom{$^{4}$}  &     $14.9^{+11.0}_{-4.0}$\negmedspace\negmedspace      &      $-54.7^{+6.3}_{-13.8}$     &      $2.0^{+1.5}_{-0.7}$                 \\
                      IRAS 16352–4721     &          core     &     $30.4$     &        \hphantom{0\,}$49.6^{+20.3}_{-23.4}$  \hphantom{$^{4}$}   &       \hphantom{1}\! $8.0^{+6.8}_{-2.9}$     &       $-40.5^{+3.2}_{-1.2}$     &      $2.0^{+1.2}_{-0.4}$  \\
                           &       outflow     &     $30.4$     &       \hphantom{0} $38.0^{+41.2}_{-22.5}$  \hphantom{$^{4}$}   &      $33.5^{+0.8}_{-0.8}$     &       $-38.0^{+6.7}_{-5.0}$     &      $3.2^{+2.1}_{-1.0}$                 \\
                           &       absorption     &     $2.73$     &      \hphantom{0}  $27.8^{+12.6}_{-18.7}$  \hphantom{$^{4}$}   &       \hphantom{1}\! $2.0^{+0.5}_{-0.5}$     &       $-39.1^{+8.2}_{-2.0}$     &      $3.1^{+3.2}_{-1.1}$                 \\
                      IRAS 16547–4247     &          core     &     $28.9$     &      \hphantom{0}  $48.5^{+14.3}_{-31.3}$  \hphantom{$^{4}$}   &       \hphantom{1}\! $8.0^{+4.4}_{-2.9}$     &       $-30.6^{+3.3}_{-2.2}$     &      $1.8^{+0.6}_{-0.4}$  \\
                           &       outflow     &     $28.9$     &      \hphantom{0}  $30.7^{+56.1}_{-15.3}$   \hphantom{$^{4}$}  &      $19.4^{+6.7}_{-6.7}$     &      $-26.9^{+8.7}_{-11.7}$     &      $5.7^{+2.6}_{-1.8}$                 \\
                           &       absorption     &     $2.73$     &      \hphantom{0}   $10.6^{+11.5}_{-4.1}$   \hphantom{$^{4}$}  &      \hphantom{1}\!  $2.4^{+0.9}_{-0.7}$     &       $-29.0^{+9.0}_{-2.8}$     &      $3.2^{+3.6}_{-1.2}$                 \\
                      G351.581–00.352     &       absorption     &     $2.73$     &       \hphantom{0}  $17.9^{+6.7}_{-11.0}$   \hphantom{$^{4}$}  &       \hphantom{1}\! $6.1^{+1.6}_{-1.1}$     &      $-100.4^{+3.0}_{-6.2}$     &      $2.9^{+2.8}_{-0.7}$                 \\
                           &       absorption     &     $2.73$     &         \hphantom{0} $12.3^{+4.5}_{-8.8}$  \hphantom{$^{44}$}   &      \hphantom{1}\!  $3.8^{+1.3}_{-1.1}$     &       $-97.2^{+1.5}_{-3.1}$     &     $5.3^{+2.9}_{-1.6}$                 \\
                           &       absorption     &     $2.73$     &         \hphantom{0} $18.2^{+5.8}_{-6.2}$   \hphantom{$^{44}$}  &      \hphantom{1}\!  $2.2^{+0.6}_{-0.5}$     &       $-94.2^{+0.3}_{-0.3}$     &      $1.1^{+0.3}_{-0.3}$                 \\
                           
                            Sgr B2 (M)     &       absorption     &     $2.73$     &       \hphantom{0} $90.0^{+22.5}_{-25.6}$   \hphantom{$^{4}$}  &      \hphantom{1}\!  $6.3^{+0.7}_{-0.6}$     &        $66.9^{+0.5}_{-0.4}$     &     $11.8^{+1.5}_{-1.5}$                 \\
                            
                                    &       absorption     &     $2.73$     &       \,$118.0^{+37.7}_{-23.9}$  \hphantom{$^{4}$}   &       \hphantom{1}\! $7.9^{+0.9}_{-0.7}$     &        $62.0^{+0.8}_{-0.7}$     &      $4.8^{+0.7}_{-0.4}$                 \\
                                    
                                    &       absorption     &     $2.73$     &          \hphantom{00} $8.5^{+3.9}_{-2.7}$   \hphantom{$^{44}$}  &      \hphantom{1}\!  $2.8^{+1.0}_{-0.3}$     &        $73.7^{+0.4}_{-5.3}$     &     $18.1^{+2.0}_{-2.4}$                 \\
                                    
                  HGAL 000.546–00.852     &          core     &     $27.2$     &        \hphantom{0} $25.1^{+22.2}_{-7.3}$   \hphantom{$^{4}$}  &      \hphantom{1}\!  $2.9^{+3.7}_{-0.4}$     &        $16.6^{+1.3}_{-0.4}$     &      $4.4^{+7.0}_{-1.3}$  \\
                          &       absorption     &     $2.73$     &       \hphantom{0} $34.0^{+13.3}_{-11.3}$   \hphantom{$^{4}$}  &      \hphantom{1}\!  $2.4^{+0.5}_{-0.3}$     &        $20.9^{+3.9}_{-0.1}$     &      $6.8^{+3.4}_{-1.8}$                 \\
                      G010.472+00.027     &          core     &     $25.1$     &       \hphantom{0} $63.2^{+78.3}_{-26.3}$   \hphantom{$^{4}$}  &       \hphantom{1}\! $3.0^{+0.8}_{-0.8}$     &                      $65.4^{+0.5}_{-0.5}$     &      $2.0^{+2.9}_{-0.7}$  \\
                           &       outflow     &     $25.1$     &         \hphantom{00}  $2.4^{+1.1}_{-1.2}$   \hphantom{$^{44}$}  &       \hphantom{1}\! $3.3^{+1.7}_{-0.9}$     &        $72.8^{+0.9}_{-0.6}$     &      $3.2^{+4.2}_{-0.7}$                 \\
                           &       absorption     &     $2.73$     &        \hphantom{0}  $13.4^{+5.1}_{-3.1}$   \hphantom{$^{44}$}  &      $17.4^{+3.3}_{-2.7}$     &        $64.6^{+1.0}_{-3.0}$     &      $1.7^{+0.5}_{-0.4}$                 \\
              W31~C      &          core     &     $31.0$     &        \hphantom{0\,}$45.1^{+20.9}_{-35.3}$   \hphantom{$^{\,\,}$}  &      \hphantom{1}\!  $4.7^{+2.3}_{-1.3}$     &        $-3.5^{+5.2}_{-8.5}$     &      $0.7^{+0.5}_{-0.2}$  \\
              (G010.624–00.384)     &       outflow     &     $31.0$     &        \hphantom{0} $22.0^{+8.4}_{-16.0}$  \hphantom{$^{4}$}   &      \hphantom{1}\!  $7.5^{+3.4}_{-1.8}$     &       $-8.0^{+2.9}_{-11.0}$     &      $2.3^{+1.9}_{-1.5}$                 \\
                     &       absorption     &     $2.73$     &         \hphantom{00}  $4.3^{+3.0}_{-1.8}$  \hphantom{$^{44}$}   &       \hphantom{1}\! $2.5^{+0.7}_{-0.7}$     &        $-2.0^{+5.4}_{-6.6}$     &      $2.1^{+1.0}_{-0.7}$                 \\
                     &       absorption     &     $2.73$     &         \hphantom{00}  $8.6^{+3.4}_{-4.9}$   \hphantom{$^{44}$}  &      \hphantom{1}\!  $2.3^{+0.7}_{-0.6}$     &         $1.2^{+9.3}_{-2.9}$     &      $2.1^{+1.0}_{-0.7}$                 \\
                      G019.609–00.234     &          core     &     $29.6$     &        \hphantom{00}  $9.5^{+15.5}_{-3.9}$  \hphantom{$^{4}$}   &      $14.5^{+8.7}_{-7.2}$     &      $41.7^{+17.3}_{-16.3}$     &     ---  \\
                               &       outflow     &     $29.6$     &        \hphantom{00}   $4.5^{+4.9}_{-1.2}$  \hphantom{$^{44}$}   &      \hphantom{1}\!  $7.9^{+8.9}_{-3.8}$     &      $27.7^{+10.6}_{-18.8}$     &      $2.3^{+2.1}_{-0.7}$                 \\
                               &       absorption     &     $2.73$     &       \hphantom{0}   $12.6^{+6.9}_{-7.0}$   \hphantom{$^{44}$}  &      \hphantom{1}\!  $2.9^{+1.4}_{-0.8}$     &        $44.7^{+8.0}_{-5.6}$     &      $2.3^{+1.7}_{-0.9}$                 \\
                      G031.412+00.307     &          core     &     $22.0$     &      \hphantom{0}  $23.5^{+12.8}_{-17.5}$  \hphantom{$^{4}$}   &       \hphantom{1}\! $3.7^{+1.4}_{-1.7}$     &       $97.3^{+4.2}_{-11.2}$     &      $1.9^{+1.6}_{-0.7}$  \\
                               &       outflow     &     $22.0$     &      \hphantom{0}  $30.1^{+14.3}_{-20.7}$   \hphantom{$^{4}$}  &      $25.5^{+4.7}_{-8.6}$     &       $97.4^{+8.7}_{-10.7}$     &      $2.3^{+1.2}_{-0.8}$                 \\
                               &       absorption     &     $2.73$     &      \hphantom{0}   $10.0^{+10.1}_{-4.3}$   \hphantom{$^{4}$}  &      \hphantom{1}\!  $2.3^{+1.1}_{-0.5}$     &       $99.4^{+10.3}_{-6.1}$     &      $2.0^{+2.0}_{-0.7}$                 \\
                             W43-MM1     &       outflow     &     $25.1$     &      \hphantom{0}  $16.4^{+12.4}_{-10.6}$   \hphantom{$^{4}$}  &       \hphantom{1}\! $8.4^{+4.7}_{-0.2}$     &       $106.1^{+4.4}_{-1.3}$     &     ---                 \\
                                     &       absorption     &     $2.73$     &        \hphantom{00}  $6.4^{+18.7}_{-0.4}$  \hphantom{$^{4}$}   &      \hphantom{1}\!  $2.6^{+2.1}_{-1.1}$     &       $102.6^{+4.8}_{-0.2}$     &      $7.5^{+2.2}_{-3.0}$                 \\
                                     &       absorption     &     $2.73$     &      \hphantom{0}   $18.7^{+8.8}_{-11.5}$  \hphantom{$^{4}$}   &      \hphantom{1}\!  $4.2^{+3.0}_{-1.6}$     &        $98.7^{+1.2}_{-0.8}$     &      $2.8^{+4.1}_{-0.5}$                 \\
                      G032.797+00.191     &          core     &     $14.6$     &     $241.0^{+453.0}_{-105.0}$     &       \hphantom{1}\! $7.1^{+8.1}_{-3.3}$     &       $14.9^{+5.9}_{-10.0}$     &      $1.3^{+2.3}_{-0.4}$  \\
                               &       outflow     &     $14.6$     &       \hphantom{0} $26.6^{+25.4}_{-18.9}$   \hphantom{$^{4}$}  &      \hphantom{1}\!  $5.1^{+5.0}_{-1.7}$     &        $5.4^{+7.9}_{-11.6}$     &      $2.7^{+1.7}_{-1.2}$                 \\
                               &       absorption     &     $2.73$     &      \hphantom{\,}  $44.9^{+289.0}_{-0.1}$     &      \hphantom{1}\!  $4.0^{+2.5}_{-1.6}$     &        $14.6^{+9.9}_{-8.6}$     &      $2.0^{+2.0}_{-0.9}$                 \\
                      G034.258+00.154     &          core     &     $29.2$     &       $243.0^{+97.1}_{-58.4}$  \hphantom{\,}   &      \hphantom{1}\!  $3.5^{+0.5}_{-0.5}$     &        $57.6^{+7.8}_{-0.7}$     &      $2.6^{+0.5}_{-0.7}$  \\
                               &       outflow     &     $29.2$     &      \hphantom{0}  $87.7^{+36.4}_{-35.6}$  \hphantom{$^{4}$}   &     \hphantom{1}\!   $5.5^{+1.3}_{-0.4}$     &        $63.8^{+3.0}_{-0.3}$     &      $3.8^{+0.7}_{-0.4}$                 \\
                               &       absorption     &     $2.73$     &       \hphantom{0} $84.4^{+16.8}_{-29.1}$   \hphantom{$^{4}$}  &      \hphantom{1}\!  $4.3^{+1.1}_{-0.3}$     &        $62.1^{+5.6}_{-0.3}$     &      $2.3^{+0.4}_{-0.3}$                 \\
                                W49~N     &          core     &     $33.3$     &      \hphantom{0}  $98.2^{+15.8}_{-28.9}$  \hphantom{$^{4}$}   &      $14.0^{+1.4}_{-0.9}$     &         $4.0^{+7.4}_{-9.2}$     &      $5.4^{+0.8}_{-0.8}$  \\
                                         &          core     &     $33.3$     &      $368.0^{+130.0}_{-96.1}$     &      $16.2^{+0.0}_{-1.5}$     &        $13.2^{+2.3}_{-2.4}$     &      $2.0^{+1.2}_{-0.1}$  \\
                                         &       absorption     &     $2.73$     &       \hphantom{00}    $1.9^{+1.4}_{-0.3}$  \hphantom{$^{44}$}   &     \hphantom{1}\!   $3.2^{+0.8}_{-0.7}$     &        $10.7^{+1.2}_{-3.1}$     &      $3.5^{+0.2}_{-0.2}$                 \\
                                         &       absorption     &     $2.73$     &       \hphantom{0}   $38.8^{+0.3}_{-3.9}$  \hphantom{$^{44}$}   &      $11.2^{+0.7}_{-0.4}$     &         $8.6^{+3.7}_{-4.3}$     &      $3.5^{+0.2}_{-0.2}$                 \\
            \hline
        \end{tabular}
        \tablefoot{ For sources known under multiple designations, the alternative designation not used in this work is provided in italics below the source name. The errors correspond to 1$\sigma$ confidence intervals. $^{(a)}$ XCLASS assumes LTE, such that $T_\mathrm{rot}=T_\mathrm{ex}=T_\mathrm{kin}$, further we assume $T_\mathrm{rot}=T_\mathrm{dust}$ from \citet{schuller2009atlasgal}, $^{(b)}$ $R_1 = \frac{N(\mathrm{H^{35}Cl})}{N(\mathrm{H^{37}Cl})}$.}
    
    \end{table*}

    \twocolumn
    \FloatBarrier
    \section{Impact of fixing \texorpdfstring{$T_\mathrm{rot}$}{T_rot} to \texorpdfstring{$T_\mathrm{dust}$}{T_dust} on column density}
    \label{sec:appendix-fixing-Trot}
    Because the atmosphere at Llano de Chajnantor is opaque to higher rotational transitions of HCl (e.g., $J=2$$-$$1$ at 1251\,GHz), only the HCl\,(1$-$0) transition is available for ground-based observations. As a result, the rotational temperature cannot be independently constrained, and the degeneracy between column density and $T_\mathrm{rot}$ must be broken by fixing one of the two parameters. We therefore adopt $T_\mathrm{rot}=T_\mathrm{dust}$, using the dust temperatures from SED fitting to ATLASGAL and Hi-GAL data \citep{schuller2009atlasgal, elia2021hi} (see Sec.~\ref{sec:spec-line-analysis-xclass}).  
    
    \begin{figure}[htbp]
        \centering
        \includegraphics[width=0.5\textwidth]{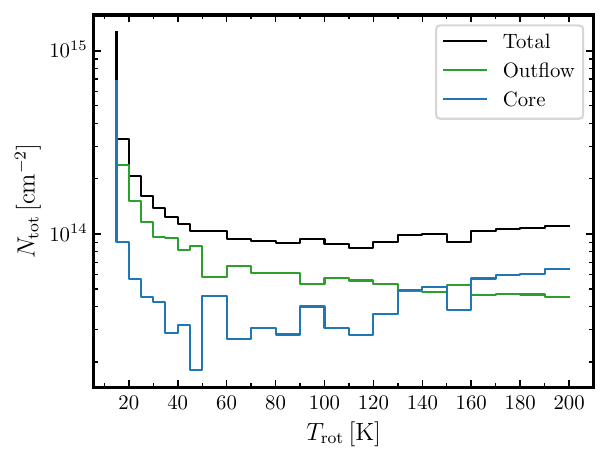}
        \caption[Effects of a fixed excitation temperature on the total column density]{Inspection of the degeneracy between column density and rotational temperature. This example is carried out on the HCl spectrum toward HGAL~{285.264$-$00.049}, observed in emission. Two components are fitted to model contributions from the core and outflow. In each run, all XCLASS/\texttt{myXCLASSFit} fit parameters are fixed, except for the column density, which is allowed to vary with the different fixed rotational temperatures.}
        \label{fig:degeneracy}
    \end{figure}
    
    To evaluate the effect of this assumption, we repeated the fits to the HCl\,(1$-$0) spectrum toward {HGAL285.264$-$00.049} for a range of fixed $T_\mathrm{rot}$ values, while keeping all other parameters constant at the resulting fit values from Table~\ref{tab:xclass-results-emission}. The resulting column densities are shown in Fig.~\ref{fig:degeneracy}. For temperatures between 20 and 50\,K, typical for massive star-forming clumps, the total $N_\mathrm{tot}$ varies by less than a factor of three. For lower temperatures ($T_\mathrm{rot}\lesssim20$ K), the fitted column densities rise steeply, whereas for $T_\mathrm{rot}\gtrsim50$\,K, the values become largely stable. The column densities of both components are left open to fit simultaneously, hence the column density of the core or outflow may be favoured in one fit compared to another, explaining the scatter over the temperatures.
    Yet, a small positive correlation of column density with rotational temperature is observed beyond $120\,$K for the narrow core component, while a small negative trend is observed for the outflow.

    The resulting column densities indicate that by fixing the rotational temperature for the usually cold cores, a source of error on the resulting column density is introduced. However, neglecting temperatures of $\leq\!15\,$K, this error is much less than an order of magnitude. In the case of the (usually hot) outflows, it indicates an overestimation of $\sim\!\!10\,\%$. Lastly, it should be noted again that the actual dust temperature of a source does not have to correspond to the rotational temperature of its HCl$\,(1$$-$$0)$ line.

    Most sources in our sample have dust temperatures around 30 K, for which the corresponding uncertainty in $N_\mathrm{tot}$ is well within the overall calibration and modelling errors. Although the true rotational temperature of HCl\,(1$-$0) does not necessarily equal the dust temperature, neglecting unrealistically low temperatures ($\leq15$ K) limits the resulting error to within a factor of a few, which is small compared to other sources of uncertainty in the analysis. For instance, the dust temperature of some of the sources with complex HCl absorption line profiles are as low as $14.6\,$K. Consequently, the error by overestimation can be more severe, but will still be small compared to ambiguities arising from the superposition of emission and absorption lines.

\end{appendix}

\end{document}